\documentclass[published]{JHEP3}
\skip\footins = 1\bigskipamount plus 2pt minus 4pt

\usepackage{epsfig}

\newcommand{\SO}{\mathop{\rm SO}\nolimits}
\newcommand{\ISO}{\mathop{\rm ISO}\nolimits}
\newcommand{\sgn}{\mathop{\rm sgn}\nolimits}

\title{Cosmological spacetimes from negative tension brane backgrounds}

\author{Cliff P. Burgess\\
Physics Department, McGill University, 3600 University Street\\
Montr\'eal, Qu\'ebec H3A 2T8, Canada\\
E-mail: \email{cliff@physics.mcgill.ca}}

\author{Fernando Quevedo and Ivonne Zavala C.\\
Centre for Mathematical Sciences, DAMTP, University of Cambridge\\ 
Cambridge CB3 0WA UK\\
E-mail: \email{f.quevedo@damtp.cam.ac.uk}, \email{eiz20@damtp.cam.ac.uk}}

\author{Soo-Jong Rey\\ 
School of Physics and Center for Theoretical Physics\\
Seoul National University, Seoul 151-747 Korea\\
E-mail: \email{sjrey@phya.snu.ac.kr}}

\author{Gianmassimo Tasinato\\
SISSA, Via Beirut 2-4, 34013 Trieste and
INFN, Sezione di Trieste, Italy\\
E-mail: \email{tasinato@he.sissa.it}}

\abstract{We identify a time-dependent class of metrics with
potential applications to cosmology, which emerge from
negative-tension branes. The cosmology is based on a general
class of solutions to Einstein-dilaton-Maxwell theory, presented
in {\tt hep-th/0106120}. We argue that solutions with hyperbolic
or planar symmetry  describe the gravitational interactions of a
pair of negative-tension $q$-branes. These spacetimes are static
near each brane, but become time-dependent and expanding at late
epoch --- in some cases asymptotically approaching flat space. We
interpret this expansion as being the spacetime's response to the
branes' presence. The time-dependent regions provide explicit
examples of cosmological spacetimes with past horizons and no
past naked singularities. The past horizons can be interpreted
as  S-branes. We prove that the singularities in the static
regions are repulsive to time-like geodesics, extract a
cosmological ``bounce'' interpretation, compute the explicit
charge and tension of the branes, analyse the classical stability
of the solution (in particular of the horizons) and study particle
production, deriving a general expression for Hawking's
temperature as well as the associated entropy.}


\received{July 30, 2002}

\accepted{October 10, 2002}

\JHEP{10(2002)028}

\begin{document}

\section{Introduction}

\rightline{\it The shortest path between two truths in the real
domain}
\rightline{{\it passes through the complex domain.} --- J.~Hadamard}

\subsection{Motivation}
Recently, there has been considerable interest in the dynamics of
brane interactions. The interest was motivated partly by the
insights which static brane configurations have already given to
long-standing low-energy issues like the hierarchy problem, and
partly by the potential application of brane
collision/annihilation processes to the cosmology of the very
early Universe~\cite{BBInflation,Ekpyrosis}. These developments
have proceeded in parallel with renewed effort toward
understanding string theory on time-dependent, cosmological
backgrounds~\cite{andre}--\cite{horowitz}.

A remarkable feature, which has emerged from studies of brane
physics, is the existence of physically sensible objects with
negative tension --- a prime example being
orientifolds~\cite{NegSense}. These objects are expected to bear
important implications for cosmology. For example, some
negative-tension objects do not satisfy the standard
positive-energy conditions which underlie the singularity
theorems. As such, they may open up qualitatively new kinds of
behaviour for the very early Universe. Negative-tension objects
also admit the possibility of zero-tension objects, whose
existence may shed light on the origin of the currently small size
of the cosmological constant~\cite{BMQ}. For instance, an
effectively tensionless 3-brane can be constructed by wrapping a
combination of Dirichlet and orientifold 5-branes on two
small-sized internal dimensions.

It then behooves us to construct a  cosmology built out of
objects having negative- or zero-tension. In so doing, it is
imperative to understand first the large-scale gravitational
fields produced by these objects. In this paper, we take a step
towards improving our understanding of these objects by providing
a class of simple space-times which describe gravitating
negative-tension objects, based on the solutions of~\cite{gqtz},
which describe cosmological spacetimes with a horizon, but with
singularities only in the static region of the full spacetime.
We  will see that these singularities correspond to
negative-tension branes of opposite charge which is consistent
with the interpretation given in~\cite{cornalba} of similar
geometries in terms of orientifold planes (although as we will
see the singularities in our metrics are not necessarily
orientifold planes). We believe these space-times are useful for
developing intuition concerning such objects, as they are no more
complicated to analyze than the well-known Schwarzschild
black-hole. The space-times we discuss in this paper enjoy the
following properties:
\begin{itemize}

\item They are classical solutions to the combined field equations
involving dilaton, metric, $(q+1)$-form $(q \ge 0)$ antisymmetric
tensor fields. For particular choices of coupling parameters, they
are classical solutions to bosonic field equations of supergravity
and string effective field theory at low-energy.
\item They describe field configurations of a pair of $q$-branes
carrying mutually opposite $q$-form charge and equal but \emph{negative} tension. These $q$-branes constitute \emph{time-like}
singularities of the space-time metric which are separated from
one another by an infinite proper distance.
\end{itemize}

We would like to interpret these space-times, containing a pair of
negative-tension objects, as counterparts of (higher-dimensional
versions of) the so-called C-metric space-time~\cite{CMetric} (see
also their dilatonic generalizations~\cite{CMetricD}), describing
the behavior of a pair of charged particles in acceleration due to
nodal singularities~\cite{CMetricInt}. Unlike the generic C-metric
solutions, however, the ones we present here have no nodal
singularity, which we interpret as meaning that no additional
stress-energy (like the ``rods and ropes'' of the C-metric) is
required in order to induce the negative-tension branes to move
along their given world-surfaces.

The space-time is time-independent in the immediate vicinity of
each brane. The static nature of the space-time metric may be
understood as a consequence of Birkhoff's and Israel's theorems
for negative- and zero-tension objects. By contrast, part of the
space-time which lies to the future of both branes is
time-dependent. The boundary between the two regions ---
time-independent versus time-dependent regions --- is a horizon of
the space-time. Curiously, the time-dependent part of the
space-time resembles that of recently discussed S-brane
configurations~\cite{stromin}.

\subsection{Negative tension versus stability}
As many of the unusual features of these space-times are traceable
to the fact that the source carries negative tension, it is worth
recalling why such branes are believed to make
sense~\cite{NegSense} --- and potentially to be
virtues~\cite{NegVirtue} --- in the low-energy world.
Traditionally, negative-tension (and negative-mass) objects were
considered pathological on the following grounds. Consider the
world-volume action of a single $q$-brane, which has the form
\begin{equation}
S_b = - \, T \int d^{q+1}y  \sqrt{-\det\gamma} + \cdots\,,
\end{equation}
where $T$ denotes the brane's tension,\footnote{Here, we tacitly
assume that the object moves relativistically so that the energy
density $\rho$ (as measured per unit $q$-dimensional volume)
equals to the tension $T$.} $\gamma_{ab} = g_{MN} \partial_a x^M
\,\partial_b x^N$ refers to the metric induced on brane's
world-volume by the space-time metric $g_{MN}(x)$, and the
ellipses represent contributions of other low-energy modes of the
brane dynamics. If the embedding of the brane were free to
fluctuate about some fixed value, $x_0^M$, in the ambient
space-time, then $x^M = x_0^M + \xi^M$ and $\xi^M$ is a \emph{dynamical} variable. A negative-definite value of the tension,
$T<0$, poses a problem since it implies a negative-definite
kinetic energy --- and hence an instability --- for the fluctuation
$\xi^M$~\cite{OtherStab}. This being so, one always assumes that
the tension $T$ of a dynamical object is positive-definite.

The explicit construction of sensible negative-tension objects
such as the orientifolds within string theory hints how the
aforementioned instability and no-go argument are avoidable.
Specifically, the argument does not apply in the instances of the
space-time studied in this paper, simply because these objects are
not free to move in the ambient space-time. Rather,
negative-tension branes are arranged to be localized at special
points, such as orbifold fixed points or space-time boundaries,
and hence do not carry dynamical variables such as $\xi^M$,
causing an instability as the tension $T$ becomes
negative-valued.\pagebreak[3] The immobility is consistent with
the equations of motion because it is the equation of motion for
the missing dynamical variables $\xi^M$ which would have required
the brane's centre-of-mass to follow a geodesic trajectory (if
the brane were neutral).

We believe the immobility of the negative-tension branes helps
explaining several otherwise puzzling features of the spacetime we
describe in this paper. For instance, as will be shown later, the
source branes do not follow geodesics in the spacetime, even when the
branes are arranged not to carry any electric charge. On the other
hand, despite not following the geodesics, the spacetime contains no
nodal singularity. The situation is unlike what arises with the
C-metric solution, where the nodal singularities are interpreted as
consequences of the external stress-energy which is required to force
the sources to move along their non-geodesic trajectories. This kind
of stress-energy is not required for negative-tension objects since,
by construction, they are not required to move along geodesics in any
case. The immobility of these objects might also help explaining why
the late-time regions of the metric are time-dependent.\footnote{We
are largely concerned with \emph{classical} aspects of the
negative-tension objects. The stability issue creeps out again once
quantum effects such as pair-creation/annihilation of these objects
are taken into account. We discuss this issue further in later
sections.}

\subsection{Outline}
This paper is structured in the following way. First, in
section~\ref{Sec2}, we review our solutions, with a particularly
simple Schwarzschild-like example, for which a generalization of
the Birkhoff and the Israel theorems~\cite{israelthm} to
negative-tension objects applies. We continue in
section~\ref{Sec3} to a much more general class of solutions. We
also show in this section how special cases of these solutions
reduce to various configurations which have been considered
elsewhere in the literature. Section~\ref{Sec4} supports the
interpretation in terms of negative-tension sources in two ways.
First, the conserved charges which are carried by the source
branes are computed using the curved-space generalizations of
Noether's theorem. Second, the response of a test particle to the
gravitational field is examined through the study of time-like
and null geodesics. Section~\ref{scbounce} describes how the
throat between the two cosmological regions can be interpreted as
a time-like bounce. Section~\ref{Sec5} investigates small
fluctuations about the solutions, with evidence presented for the
instability of some of their remote-past features. We believe the
late-time metric to be stable, and we regard the calculations of
this section as a first step towards a more comprehensive
stability analysis. We also discuss in this section the relevance
of the time-like singularities, and why these can make sense of
space-time's overall causal structure. In this section, we show
that a Hawking temperature can be defined, and we present
preliminary arguments that this reflects the spectrum of
particles seen by static observers. Finally, we summarize our
conclusions in section~\ref{Sec6}, where we also comment on some
future directions for research which our calculations suggest,
above all on the construction of the cosmological models.

\section{Simple solutions}\label{Sec2}

Before presenting our solutions in their most general form, we
pause here first to build intuition by describing their simplest
variant: vacuum solution to Einstein's field equation, $R_{\mu\nu}
= 0$, in four dimensions.

\subsection{Schwarzschild revisited}
We start with the well-known Schwarzschild black-hole, whose
space-time geometry is given --- in the asymptotically flat region
$r\geq 2M$ --- by:
\begin{equation}\label{uno}
ds^2_I = - \left[1-\frac{2M}{r}\right] dt^2 +
\left[1-\frac{2M}{r}\right]^{-1} dr^2 +
r^2\left(\sin^2\theta d\phi^2 +d\theta^2\right),
\end{equation}
whose constant $r$ and $t$ surface is the two-sphere $S_2$. \emph{Birkhoff's theorem} states that eq.~(\ref{uno}) is the unique
solution for representing \emph{spherically symmetric} non-rotating
black holes.\footnote{We emphasize that this theorem assumes
nothing regarding time-(in)dependence of the solution.} \emph{Israel's theorem}~\cite{israelthm} states further that
eq.~(\ref{uno}) is also the unique solution for representing \emph{static} non-rotating black holes.\footnote{ Although we describe
in detail in this section the four-dimensional case, our
discussion trivially generalizes to $2+n$ dimensions --- with
$n\ge 2$ --- through the replacement $1/r \rightarrow 1/r^{n-1}$.}

As is well known, the apparent singularity of the metric
eq.~(\ref{uno}) on the surface $r=2M$ is a coordinate artifact.
For $r<2M$, the metric goes over to that of the interior region,
for which the role of $r$ and $t$ gets interchanged, leading to a
time-dependent metric of the form:\footnote{We adopt here the
convention of always labelling the time coordinate as $t$, both
inside and outside the horizon.}
\begin{equation}\label{dos}
ds^2_\mathrm{II} = - \left[\frac{2M}{t} - 1 \right]^{-1} dt^2 +
\left[\frac{2M}{t} - 1\right] dr^2 +
t^2\left(\sin^2\theta d\phi^2 +d\theta^2\right).
\end{equation}
Note that the surface of constant $r$ and $t$ remains the same
two-sphere $S_2$. A real, \emph{space-like} singularity occurs for
$t\rightarrow 0$, which is to the future of any observer falling
into the Schwarzschild black-hole.

A particularly simple form of the solutions which are of interest in
this paper may be obtained from eq.~(\ref{dos}) by an \emph{analytic
continuation}, $\theta \rightarrow i \theta$ followed by an overall
sign change of the metric.\footnote{Equivalently we can take
$\theta\rightarrow i\theta, \phi\rightarrow i\phi, t\rightarrow ir, r
\rightarrow it, M\rightarrow iP$.} This leads to the following
time-dependent vacuum solution:
\begin{equation}
\label{tres} ds^2_{I} = - \left[1-\frac{2P}{t}\right]^{-1} \
dt^2\  +\ \left[1-\frac{2P}{t}\right]\ dr^2 +\
t^2\left(\sinh^2\theta d\phi^2 +d\theta^2\right).
\end{equation}
Note that, after the analytic continuation, the surface of
constant $r$ and $t$ has turned from the two-sphere, $S_2$, to the
hyperbolic surface, $\mathcal{H}_2$, viz. sign of the curvature
scalar is flipped from positive to negative. The metric is
explicitly time-dependent but homogeneous otherwise --- it has a
space-like Killing vector $\xi = \partial_r$ in addition to the
symmetries of the hyperbolic surface $\mathcal{H}_2$ at fixed
values of $r$ and $t$.

Equation~(\ref{tres}) is well-defined for $t > 2P$, but as is
clear from its connection with the Schwarzschild black-hole, the
degeneracy at $t = 2P$ is merely a coordinate artifact. An
extension of the metric to $t< 2P$ is given by performing the
same continuation as the one leading to eq.~(\ref{tres}):
\begin{equation}\label{cuatro}
ds^2_\mathrm{II} = - \left[\frac{2P}{r}-1\right] dt^2 +
\left[\frac{2P}{r} - 1\right]^{-1} dr^2 + r^2\left(\sinh^2\theta
d\phi^2 +d\theta^2\right).
\end{equation}
The metric in this region is static and retains the hyperbolic space
$\mathcal{H}_2$ at constant $r$ and $t$. A real, \emph{time-like}
singularity occurs for $r\rightarrow 0$, and this is the structure we
are primarily interested in this paper.

\EPSFIGURE{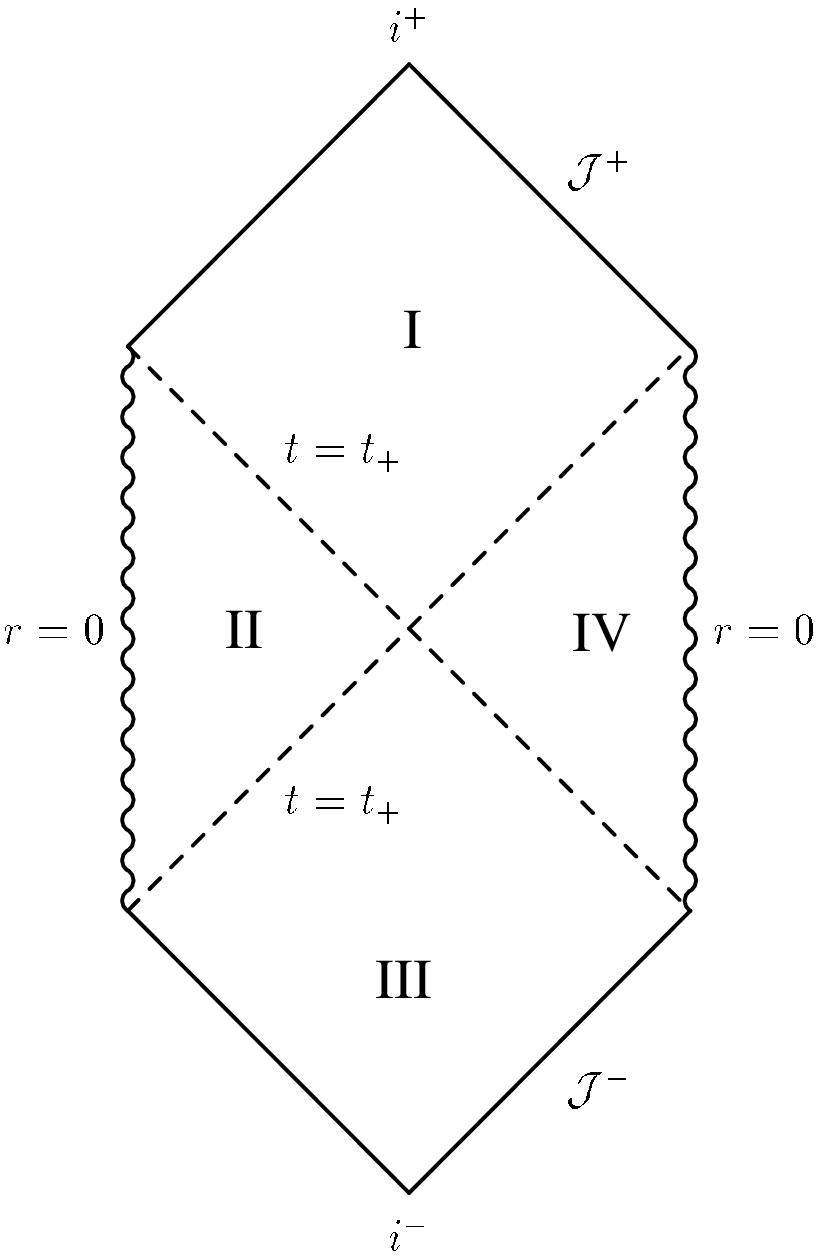, width=.43\textwidth}{Penrose diagram
for the $k=0,-1$ case. This diagram is very similar to the
Schwarzschild black hole (rotated by $\pi/2)$, but now region I
(III) is not static, but time-dependent with a Cauchy horizon
and region II (IV) is static.\label{fig1}}

Just as $r=2M$ does for the Schwarzschild black hole, the surface
$t=2P$ defines a \emph{non-compact}  horizon of the space-time
described by eqs.~(\ref{tres}) and~(\ref{cuatro}). This is most
transparently seen from the Penrose diagram of the space-time,
given in figure~\ref{fig1}. It is simply a $\pi/2$-rotation of the
Penrose diagram for the Schwarzschild space-time.

An observer in region I experiences a time-dependent, expanding
region of the space-time,  which becomes \emph{flat} as $t \rightarrow
\infty$. The observer sees no singularity in null or time-like
future, but will experience two time-like singularities in the
past. By contrast, an observer in regions II and IV experiences a
static space-time, and sees only a single time-like singularity in
the past. Observers in region III see no singularities to their
past, but have both time-like singularities in their future light
cones. On the other hand, observers at fixed values of $r,\theta$
and $\phi$ in the static regions --- including the singularities
themselves as a limiting case --- do not follow geodesics and so
experience a proper acceleration.

The above description suggests a viable interpretation for this
solution, as well as for many of the other more general ones which
we present in subsequent sections. Regions II and IV describe the
space-time external to two objects which we argue to be
negative-tension branes. These branes may also carry other
conserved charges. Region I gives the time-varying transient
gravitational fields which these branes produce at late times.
Region III similarly describes the time-reversal of this last
time-dependent process.

\looseness=1In this interpretation, the horizons, which are reminiscent of
S-branes (in a precise sense explained below)~\cite{stromin},
describe the locus of instants when observers make the transition
from having only one of the branes in their past light cone to
having them both in their past.

\subsection{Kruskal coordinates}
For understanding the overall structure of the above space-time,
it is useful to have explicit coordinates for the complete
maximally-extended space-time, whose Penrose diagram is pictured
in figure~\ref{fig1}. Starting from the geodesically incomplete
solutions eqs.~(\ref{tres}) and~(\ref{cuatro}), a convenient
choice of coordinates for the extended geometry is obtained by
using the analog of the Kruskal coordinates, defined as follows:
\begin{itemize}
\item For $t \ge r_{+}$, define
\begin{eqnarray}
v & := & \pm \left[\frac{t}{2 P}-1 \right]^{{1}/{2}}
e^{t/4 P} \cosh\left( \frac{r}{4P} \right), \nonumber\\
u & := & \left[ \frac{t}{2 P}-1 \right]^{{1}/{2}}  e^{t/4P}
\sinh \left( \frac{r}{4 P} \right),
\end{eqnarray}
where the square-root is taken to be positive, and the upper and
the lower signs correspond to region I and III, respectively.

\item For $r \le r_{+}$, define
\begin{eqnarray}
v  &:=&  \left[ 1-\frac{r}{2 P} \right]^{{1}/{2}} e^{r/4 P}
\sinh \left( \frac{t}{4 P} \right),  \nonumber\\
u & := & \mp \left[ 1-  \frac{r}{2 P} \right]^{{1}/{2}}
e^{r/4 P} \cosh\left( \frac{t}{4 P} \right),
\end{eqnarray}
where again the positive square-root is understood, and the upper
and the lower signs correspond to regions II and IV, respectively.
\end{itemize}

For the time-dependent regions, I and III, the metric in the
Kruskal coordinates \mbox{becomes}
\begin{eqnarray}
d s^{2} = \frac{16 P^{3}}{t}  e^{-{t}/{2 P}}(-d v^{2}+d u^{2}) +
t^{2} \left( \sinh^2\theta d\phi^2 +d\theta^2 \right).
\end{eqnarray}
The Penrose diagram of figure~\ref{fig1} follows from this metric
after performing a straightforward conformal transformation which
brings the asymptopia in to a finite-distance. The horizon
corresponds in these coordinates to the lines $u=\pm v$.
Similarly, in the static regions II and IV, where $r<r_{+}$, we
have the relation
\begin{eqnarray}
u^{2}-v^{2}= \left[1- \frac{r}{2P}\right] e^{r/2 P}\,.
\end{eqnarray}
The singularity, $r=0$, is then given by the hyperbola
$u^{2}=v^{2} + 1$. Using these coordinates it is straightforward
to see that the proper distance between each singularity and the
horizon is infinite, and therefore the two singularities have to
be thought to be infinitely separated.

\subsection{Alternative analytic continuations}
In addition to the one we have adopted, can one find yet another
analytic continuations of the Schwarzschild black-hole which would
also admit cosmological interpretations in the asymptotic regions?
Some time ago, Gott~\cite{gott} (see also~\cite{others}) proposed
an alternative analytic continuation: For regions I and III, his
continuation is the same as ours: $\theta \rightarrow i \theta$
followed by an overall signature change. For regions II and IV,
his differs from ours and is given simply by $\theta
\rightarrow \pi/2 + i \tau$ and $t \rightarrow i \psi$ applied to
eq.~(\ref{uno}). Thus,
\begin{eqnarray}
ds_\mathrm{I, III}^2 &=& - \left[ 1 - \frac{2 M}{t} \right]^{-1}
dt^2 +\left[1 - \frac{2 M }{ t} \right] dr^2 + t^2
\left( \sinh^2 \theta d \phi^2 + d \theta^2 \right), \nonumber \\
ds_\mathrm{II, IV}^2 &=& + \left[ 1 - \frac{2 M }{ r} \right]^{-1} dr^2 +
\left[ 1 - \frac{2 M }{ r} \right] d \psi^2 + r^2 \left(\cosh^2 \tau
d \phi^2 - d \tau^2 \right).
\label{cinco}
\end{eqnarray}
His analytic continuation was built upon totally different
physical motivations from ours, and it was interpreted as giving
rise to the space-time associated with a gravitating tachyon. The
interpretation was borne out on the ground that the $r=0$
singularity is time-like and is perpendicular to the space-like
world-line of a tachyon, much the same way as the space-like
singularity of the Schwarszchild metric is related to the
time-like world-line of a massive particle at rest.\footnote{This
interpretation was later reconsidered by Gibbons and
Rasheed~\cite{gary}.} The existence of a tachyon would signal
instability of the space-time involved, and the solution
eq.~(\ref{cinco}) in fact describes a  ``\emph{bubble of
nothing}''~\cite{horowitz,witten,gary}. Note the difference
between this solution and the solution of eq.~(\ref{tres}). In
eq.~(\ref{tres}), the metric is a warped product of a lorentzian
surface, $\Sigma_{1,1}$, parameterized by $r,t$, and a euclidean
hyperbolic space $\mathcal{H}_2$, parametrized by $\theta, \phi$.
By contrast, for the gravitating tachyon, eq.~(\ref{cinco}), the
$r, \psi$ coordinates parameterize a euclidean ``cigar''
(assuming that $\psi$ is periodic), while the $\tau$ and $\phi$
coordinates describe a two-dimensional de Sitter
space.\footnote{We thank Gary Gibbons for a discussion on these
points.}

Space-time geometries whose Penrose diagrams are similar to ours
have also been studied previously for Einstein gravity coupled to
a variety of other fields~\cite{CMetricInt,collins,martinspaper}.
In the context of string theory, an example was found
in~\cite{kounnas} on which the string dynamics is describable in
terms of non compact, two-dimensional Wess-Zumino-Novikov-Witten
(WZNW) models. More recently, a general class of brane solutions
was found~\cite{gqtz} in which some of the coordinates
parameterize a subspace with constant curvature, labelled by
$k=1,0,-1$ as for the Friedman-Robertson-Walkers (FRW) metrics. In
these solutions, the case $k=1$ represents the standard
black-brane solutions of a system consisting of
gravity/dilaton/antisymmetric-tensor fields, but the cases
$k=0,-1$ exhibit the Penrose diagram similar to the one discussed
above, viz. Figure~\ref{fig1}. We shall see that eqs.~(\ref{tres})
and~(\ref{cuatro}) furnish particular cases of the general
solutions of~\cite{gqtz}. More recently, geometries similar to
ours have been considered as orbifold cosmological
models~\cite{cornalba} and as
S-branes~\cite{stromin,gutperle,myers} (see also~\cite{andre}).
In particular, interesting cosmological consequences were drawn
in~\cite{cornalba} from space-time geometries which these authors
claim are produced by a pair of orientifolds.

\section{General solutions}\label{Sec3}

We now turn to the description of a wider class of solutions which
extend the simple considerations of section~\ref{Sec2} to various
space-time and brane's world-volume dimensions, and to a system
involving metric, dilaton, and $(q+1)$-form tensor fields --- a
system encompassing bosonic fields of diverse supergravity or
superstring theories and their compactifications. This wider class
of solutions was already obtained in~\cite{gqtz}, in which the
primary interest was generalization of the well-known black branes
of string theory to all possible signs of the curvature parameter,
$k$, of the maximally-symmetric transverse space.

\subsection{Dilaton-generalized Maxwell-Einstein solutions}\label{sec3.1}
The system we will consider is defined by the following
Einstein-frame action in $d=(n+q+2)$-dimensional space-time:
\begin{equation}\label{generalaction}
S= \int_{\mathcal{M}_d} d^{d}x \sqrt{g}
\left[\alpha R - \lambda (\nabla\phi)^2 -
\eta e^{-\sigma \phi} F_{q+2}^2 \right],
\end{equation}
where $g_{\mu \nu}, \phi, F$ denotes metric, dilaton field, and
$(q+2)$-form tensor field strength, respectively. Stability
requires the constants $\alpha, \lambda$ and $\eta$ to be
positive, and, if so, they are removable by absorbing them into
redefinition of the fields.\footnote{The canonical choices are
$\lambda=\alpha=1/2$, and $\eta=1 \slash {2(q+2)!}$ in units where
$8\pi G=1$.} It is nevertheless useful to keep them arbitrary
since this would allow to examine various reduced systems, where
each constant is taken zero (to decouple the relevant fields) or
negative (e.g.\ to reproduce E-brane solutions in the
hypothetical type-II* string theories, related to the type-II
string theories via time-like T-duality. See later.).
eq.~(\ref{generalaction}) includes supergravity, and so also
low-energy string theory, for specific choices of $d$, $\sigma$
and $q$ (for instance $d=10$, $q=1$ and~$\sigma=1)$.

The field equations obtained from eq.~(\ref{generalaction}) are
given by:
\begin{eqnarray}
\alpha G_{\mu\nu} & = & \lambda T_{\mu \nu}[\phi] +
\eta e^{- \sigma \phi} T_{\mu \nu}[F]\,, \label{einstein}\\
2\,\lambda\,\nabla^2 \phi &=& -\sigma\,\eta \, e^{-\sigma\phi} F^2\,,
\label{dilaton1}\\
\nabla_{\mu} \left( e^{-\sigma\phi} F^{\mu \cdots}\right) & = & 0\,,
\label{eqgaf}
\end{eqnarray}
where
\begin{eqnarray*}
T_{\mu \nu}[\phi] = \nabla_\mu \phi \nabla_\nu \phi
-\frac{1}{2} g_{\mu \nu} (\nabla\phi)^2\,, \qquad \mbox{and}
\qquad T_{\mu \nu}[F] = (q+2) F_\mu{}^{\cdots} F_{\nu\cdots} -
\frac{1}{2}g_{\mu\nu} F^2_{\cdots}\,.
\end{eqnarray*}
We are interested in classical solutions whose space-time geometry
take the form of an asymmetrically warped product between
$q$-dimensional flat space-time and $n$-dimensional
maximally-symmetric space, parametrized by a constant curvature $k
= 0, \pm 1$. For this ans\"atz, the solutions depend only on one
warping variable --- either $t$ or $r$ --- and ought to exhibit
isometry $\SO(1,1) \times O_{k}(n) \times \ISO(q)$, where
$O_{k}(n)$ refers to $\SO(n-1,1)$, $\ISO(n)$ or $\SO(n)$ for
$k=-1$, $0$ and $1$, respectively. The ans\"atz is motivated for
describing a flat $q$-brane propagating in
$d=(n+q+2)$-dimensional ambient space-time, where $n$-dimensional
transverse hyper-surface is a space of maximal symmetry, and
constitute an extension of Birkhoff's and Israel's theorems.

A solution satisfying these requirement is readily obtained
as~\cite{gqtz}
\begin{eqnarray}
ds^2 & = & h_-^A \left( -h_+h_-^{1-(n-1) b}dt^2 +h_+^{-1}h_-^{-1+b}
dr^2 +r^2h_-^{b} dx^2_{n,k} \right) + h_-^B dy^2_{q}\,,
\qquad \label{metric}\\
\phi &=& \frac{(n-1) \sigma b}{\Sigma^2} \ln h_- \,,\label{eq:dil}\\
F_{try_1 \dots y_q} &=& Q \epsilon_{try_1 \dots y_q} r^{-n},
\qquad \epsilon_{try_1 \dots y_q}=\pm 1 \,.\label{eq:F}
\end{eqnarray}
The notations are as follows. The metric of an $n$-dimensional
maximally symmetric space, whose Ricci scalar equals to $n(n-1)k$
for $k=0,\pm 1$, is denoted as $d x^2_{n, k}$. The harmonic
functions $h_\pm(r)$ depend on two first-integral constants,
$r_\pm$, and are given by:
\begin{equation}\label{h+}
h_+(r) = s(r) \left(1-\left(\frac{r_+}{r}\right)^{n-1} \right),
\qquad h_-(r) = \left|k-\left(\frac{r_{-}}{r}\right)^{n-1}\right|,
\end{equation}
where
\begin{equation}\label{signo}
s(r)=\sgn \left(k-\left(\frac{r_{-}}{r}\right)^{n-1}\right).
\end{equation}
The constant $Q$ is given by
\begin{equation}\label{Qdef}
Q = \left(\frac{4n(n-1)^2 \alpha \lambda
(r_+r_-)^{n-1}}{(q+2)!\,\eta\,(\alpha n \Sigma^2 +
4(n-1)\,\lambda)} \right)^{1/2},
\end{equation}
where $\Sigma$ and $b$ are constants defined in terms of
parameters of the action as
\begin{eqnarray*}
\Sigma^2 = \sigma^2 + \frac{4 \lambda}{\alpha}
\frac{q(n-1)^2}{n(n+q)}\,, \qquad b = \frac{2\alpha n
\Sigma^2}{(n-1)(\alpha n \Sigma^2+ 4(n-1)\lambda)}\,.
\end{eqnarray*}
Likewise, the exponents $A, B$ in eq.~(\ref{metric}) are defined in
terms of the same parameters as
\begin{eqnarray*}
A = - \frac{4 \lambda q (n-1)^2 b }{\alpha n (n+q) \Sigma^2}\,,
\qquad \mbox{and} \qquad B = \frac{4 \lambda (n-1)^2 b}{\alpha
(n+q) \Sigma^2} = -\frac{n}{q} A\,.
\end{eqnarray*}
The solution defined by eqs.~(\ref{metric})--(\ref{eq:F}) is
unique modulo trivial field redefinitions: $\phi \rightarrow
\phi(r) + 2 \phi_0$ and $F \rightarrow F e^{\sigma \phi_{0}}$,
which in turn can be compensated by rescaling of the space-time
coordinates and the first-integral constants, $r_\pm$.

The two first-integral constants, $r_\pm$, are related intimately
to two conserved charges associated with the solution. One of
these is the $q$-form electric charge $Q$ --- see eq.~(\ref{eq:F})
--- acting as the source of the $(q+2)$-form tensor field
strength. The electric charge is measurable from the flux integral
$\oint{^*F_{q+2}}$ over the $n$-dimensional symmetric space. The
explicit integral yields the electric charge given precisely by
eq.~(\ref{Qdef}), so is a function of the first-integral
constants. For the $q$-brane to be physically sensible, the
electric charge $Q$ ought to be real-valued. From eq.~(\ref{Qdef})
and from the stability condition $\eta>0$,\footnote{This last
conclusion does not follow for E-branes, for which $\eta$ may be
chosen negative (see later sections).} the condition renders an
inequality $(r_-r_+)^{n-1}\ge 0$. Note that $r_-=0$ if and only
if $Q=0$, and the point $r=r_-$ is potentially singular (or a
horizon) only if $k=1$. The second conserved charge is associated
with the Killing vector of the metric, and so can be understood
as a mass, in a sense which will be made explicit later.

A dual, magnetically charged solution is  obtainable from
the electrically charged solution by making the duality
transformation: $F_{q+2} \rightarrow \tilde F_n = {}^*F_{q+2}$, $\sigma
\rightarrow -\sigma$ and $q \rightarrow (d-4-q)$ in
eqs.~(\ref{metric}),~(\ref{eq:dil}) and~(\ref{eq:F}), where
$F_{q+2}$ is related to $\tilde F_n$ through the
dilaton-dependent expression:
\begin{eqnarray}
F_{q+2} = \mathrm{e}^{\sigma\phi}\epsilon_{q+2,n}\tilde F_n\,.
\end{eqnarray}

The solution presented above is expressed as a function of the
coordinate $r$. One readily finds that $r$ denotes a spatial
coordinate for $k=-1,0$ in so far as $r < r_+$. For $r > r_+$, the
harmonic function $h_+$ flips the overall sign, so the $r$ coordinate
becomes temporal. As such, we will relabel the coordinates as $r
\leftrightarrow t$ for $r> r_+$ so that $t$ labels always the time
coordinate.

Drawing lessons from the simple solution presented in
section~\ref{Sec2}, we are primarily interested in $k=-1, 0$
cases. Note that the $k=-1$ solution is obtainable from the $k=1$
solution in much the same way via the following \emph{analytic
continuation}:
\begin{eqnarray*}
t &\longrightarrow & ir\,, \qquad r\longrightarrow it\,,\qquad
\Omega_n \longrightarrow i \Omega_n\,, \qquad \mbox{and}\\
\qquad r_+ &\longrightarrow & i r_+\,, \qquad r_-^{n-1} \longrightarrow
-(i r_-)^{n-1}\,.\qquad
\end{eqnarray*}
Note that this is precisely the same as that defined the simpler,
Schwarzschild-type solution in the previous section. See footnote
6. The above procedure also suggests that we can obtain yet
another solution with $h_+=\left|k-(r_{+}/r)^{n-1}\right|\, $ and
$h_-=1-(r_{-}/r)^{n-1}$ via an alternative analytic continuation
$r_- \rightarrow i r_-$ and $r_+^{n-1} \rightarrow -(i
r_+)^{n-1}$.\footnote{The additional minus sign is required to
ensure the real-valuedness of the electric charge $Q$.} It turns
out these new solutions are singular at $r=r_-$ for generic
values of the parameters. As such, they would correspond to more
standard cosmology evolving from a past singularity. Further new
(and generically singular) solutions are also obtainable by
$T$-dualizing the above solutions with respect to the coordinate
$r$ in the time-dependent region and $t$ in the static region. In
this case the corresponding element of the metric --- $g_{rr}$ or
$g_{tt}$ --- in the string frame gets inverted and the dilaton
field is shifted accordingly (see for instance, ref.~\cite{bmq1}).

\subsection{Asymptotic and near-horizon geometries}
For foregoing discussions, we pause here to examine both the
asymptotic and the near-horizon geometries of our solution,
eqs.~(\ref{metric})--(\ref{eq:F}). As the $k=1$ case parallels to
the standard black-brane studies, we focus primarily on the
$k=0,-1$ cases. For the ease of the analysis, we adopt the \emph{isotropic coordinates}, defined by
\begin{eqnarray}
\tau^{n-1} = \left( t^{n-1} - r_+^{n-1} \right).
\end{eqnarray}
The near-horizon and the asymptotic limits then correspond to
$\tau \rightarrow 0$ and $\tau \rightarrow \infty$, respectively.

The metric eq.~(\ref{metric}) takes, in the isotropic coordinates,
the form:
\begin{eqnarray}
ds^2 = \left(\frac{H_- }{H_+}\right)^{A+b}
\left[-\frac{H_+^{2 \slash(n-1)}}{H_-} d\tau^2 +
\frac{H_-^{1-nb} }{H_+^{2 - nb}}  dr^2 \!+\!
\tau^2 H_+^{2 \slash (n-1)} dx^2_{n,k} \right] \!+\!
\left(\frac{H_- }{H_+} \right)^B dy^2_{q}\,.\qquad\quad
\label{isometric}
\end{eqnarray}
The harmonic functions $H_\pm(\tau)$ are given, for the $k=-1$
case, by
\begin{eqnarray}
H_+ = 1 + \left(\frac{r_+}{\tau}\right)^{(n-1)}\,,
\qquad H_- = H_+ + \left(\frac{r_-}{\tau}\right)^{(n-1)}\,,
\end{eqnarray}
and, for $k=0$ case, by
\begin{eqnarray}
H_+ = 1 + \left(\frac{r_+}{\tau}\right)^{(n-1)},
\qquad H_- =\left(\frac{r_-}{\tau}\right)^{(n-1)}.
\end{eqnarray}
Likewise, the dilaton field and the $(q+2)$-form tensor field
strength are given in the isotropic coordinates by
\begin{eqnarray}
\phi(\tau)=\frac{(n-1) \sigma b}{\Sigma^2} \ln{(H_+^{-1}H_-)}\,, \qquad
\mbox{and} \qquad F_{try_1\dots y_q}(\tau) = Q
\epsilon_{try_1 \dots y_q} \tau^{-n}H_+^{-n/(n-1)}\,.\qquad\ \,
\label{dil&F}
\end{eqnarray}

From eqs.~(\ref{isometric}) and~(\ref{dil&F}), we now analyze the
limiting geometries for the two cases $k=-1,0$ separately.

\paragraph{The $k=-1$ brane:}
in the asymptotic region, $\tau \rightarrow \infty$, and both
$H_+$ and $H_-$ approach the unity. That is, the asymptotic
geometry is flat:
\begin{eqnarray}
ds^2 \vert_\mathrm{asymptotic} = - d\tau^2 + dr^2 +
\tau^2 d \mathcal{H}^2_{n} +  dy^2_{q}\,,
\end{eqnarray}
where $d\mathcal{H}_n^2 = dx^2_{n,-1}$. Moreover, both the
dilaton field and the $(q+2)$-form field strength become zero:
$\phi, F \rightarrow 0$. This is very interesting as the
time-dependent regions I and III tend asymptotically to a vacuum
state corresponding to (a patch of) flat space-time, both in the
asymptotic past and future infinity. 

In case the system under consideration is the bosonic part of a
supersymmetric theory, the asymptotic region could constitute a
supersymmetric vacuum. For instance, as the asymptotic geometries
are flat, in- and out-states might be defined naturally having
anywhere up to the maximal number of unbroken supersymmetries.
Clearly, cosmologies with asymptotic supersymmetry could have
many interesting features.

In the near-horizon region, $\tau \rightarrow 0$, and the harmonic
functions are reduced to
\begin{eqnarray*}
H_+ \longrightarrow \left(\frac{ r_+ }{\tau} \right)^{n-1}, \qquad
\mbox{and} \qquad H_- \longrightarrow \left(\frac{\overline{r}}{\tau}
\right)^{n-1},
\end{eqnarray*}
where $\overline{r}^{n-1} := (r_-^{n-1} + r_+^{n-1})$. For
simplicity, consider a particular choice of the parameters so
that $r_-=r_+$ --- the result does not change if they are
different. The metric then behaves~as
\begin{eqnarray}
ds^2\vert_\mathrm{near-horizon} = -d \tilde t^2 +
\left(\frac{\tilde t}{r_+}\right)^2 dr^2
+ r_+^2  d\mathcal{H}^2_{n} +  dy^2_{q}\,,
\end{eqnarray}
where unimportant numerical factors are absorbed by rescaling
coordinate variables, and the time coordinate is newly defined as
$\tilde t:=\tau^{(n-1)/2} r_+^{(n-3)/2} 2^{(1-A-b)/2}$. Note that
the near-horizon geometry does not depend explicitly on the
dimension $n$ of the transverse space. Moreover, the dilaton
field and the $(q+2)$-form field strength tend to constants in
this~limit.

\FIGURE{{\epsfig{file=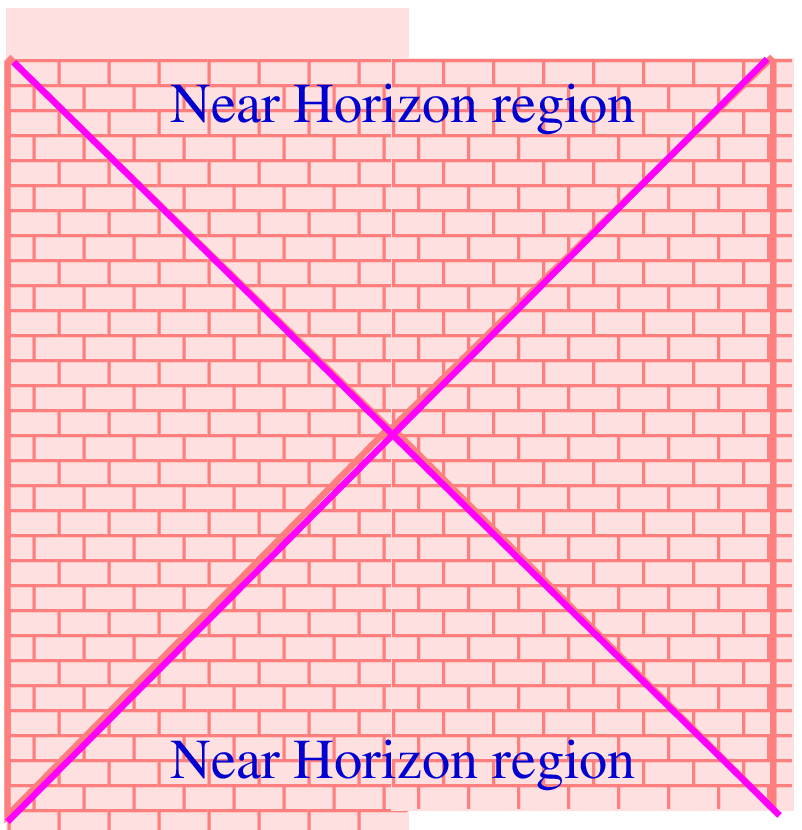, width=.35\textwidth}}%
\caption{Penrose diagram for the near horizon limit. It is simply the
Milne Universe, where the shaded zone is excluded and an apparent
singularity sits at $\tilde t=0$.}\label{fig2}}

We thus find that the near-horizon geometry of the
time-dependent regions I and III $(t > r_+)$ is described by the
direct product of a two-dimensional Milne Universe with
coordinates $\tilde{t}$ and $r$, an $n$-dimensional hyperbolic
space with coordinates $x_{n}$, and a $q$-dimensional flat space
with coordinates $y_{q}$. In the near-horizon geometry, the
Penrose diagram of figure~\ref{fig1} goes over to that of the
Milne Universe, illustrated in figure~\ref{fig2}. The apparent
singularity at $\tilde{t}=0$ is harmless, as it corresponds to a
regular point at the horizon.

Alternatively, the near-horizon limit can be taken from the static
interior regions --- regions II and IV of figure~\ref{fig1}. In
this case, we find that the two-dimensional spacetime with
coordinates $r,t$ is reduced to Rindler spacetime --- the shaded
region in figure~\ref{fig2}.

\paragraph{The $k=0$ brane:}

the $k=0$ branes exhibit several marked differences from the $k =
\pm1$ ones. The main difference is in the asymptotic geometry,
which in this case does not become flat as $\tau \rightarrow
\infty$. In particular, in this limit, the coefficient of
$d\mathcal{E}^2_n = dx_{n,0}^2$ goes to zero and the dilaton
field runs logarithmically to $\phi = -\infty$.

The result for the metric in the near-horizon limit -again taking
$r_- = r_+$ for simplicity --- is:
\begin{eqnarray}
ds^2 \vert_\mathrm{near-horizon} = -d\tilde t^2 +
\left(\frac{\tilde t}{r_+}\right)^2 dr^2
+ r_+^2  d\mathcal{E}^2_{n} + dy^2_{q}\,.
\end{eqnarray}
As might be expected starting from the original causal structure,
the geometry is again a direct product of a two-dimensional Milne
Universe, an $n$-dimensional flat space and a $q$-dimensional
flat space. The Milne Universe geometry of the near-horizon
region seems to be quite generic for all these solutions. Note,
however, that, in this case, the near horizon geometry is an
exactly flat space-time, in contrast to the $k=-1$ brane.

\subsection{Special cases}
Recently, variants of the supergravity/superstring brane solutions
were considered in the literature. We now pause to show that they
are nothing but particular cases of our brane solution
eqs.~(\ref{metric})--(\ref{eq:F}).

\paragraph{Black-branes: $k=+1$}

the $k=1$ branes were studied extensively in the literature, and
was interpreted as a black $q$-brane~\cite{strominger91}. For the
special case $q=0$ and in the absence of the dilaton field
coupling, the solution reduces (as it should) to a
higher-dimensional version of the Reissner-Nordstr\"om black-hole.
It is possible to define the Arnowitt-Deser-Misner (ADM) mass for
the black $q$-brane, and turns out to be given by $M= r_+^{n-1} +
r_-^{n-1}[1-(n-\sigma^2/\Sigma^2)b]$.

It was known that, in the case the dilaton coupling is non zero,
that the surface $r=r_+$ is a coordinate singularity corresponding
to an event horizon, while the surface $r=r_-$ is a \emph{bona
fide} null singularity of the scalar curvature. This singularity
is formed when the unstable inner horizon of the
Reissner-Nordstr\"om black-hole is perturbed and made singular by
coupling the black-hole to the dilaton field. Only if $r_-< r_+$,
is this singularity covered by the horizon, otherwise it is
naked. The point $r=0$ is also singular in the usual~sense.

Note that, in  eq.~(\ref{metric}), the region beyond the null
singularity at $r=r_-$ is well defined.  As is known, an extremal
black-brane is obtained with the choice $r_{+}=r_{-}$, which also
promotes the symmetry of the space-time to $\SO(n+1) \times
\SO(q,1)$. The solution then corresponds to the field due to a
D$q$-brane of string theory, for which the horizon becomes
singular.\footnote{Of course, this singularity is outside the
domain of validity of the low-energy string effective field
theory.} An exception to the above statement, for which the
solution \emph{is} well-defined and the dilaton field well-behaved,
is the case $q=3$.

\paragraph{S-branes: $k=0,-1$:}

consider next the $k=0,-1$ branes. These resemble the analytic
continuation of the Schwarzschild black-hole considered in
section~\ref{Sec2}, whose Penrose diagram is given by
figure~\ref{fig1}. First of all, note that, for fixed $r$ and
$t$, the sign of the harmonic function $h_+$ flips as one changes
from $k=+1$ to $k=-1,0$. This implies that roles played by $r$
and $t$ coordinates are interchanged --- see
eqs.~(\ref{metric}),~(\ref{h+}) and~(\ref{signo}) --- and the
metric is time-dependent. In this case, the point $t=r_-$ is a
regular point, while $t=r_+$ is a \emph{past Cauchy horizon}. See
figure~\ref{fig1}. The point $r=0$ is now a time-like
singularity, which is behind the past horizon of the future
time-dependent region I. Unlike the case of the
Schwarzschild-type solution, this singularity can be avoided by
future-directed time-like curves in the region between the horizon
and the singularity. See figure~\ref{fig3}.

A simple analysis indicates that, for $k=-1,0$, there is no real-
and positive-valued choice for $r_+$ and $r_-$, for which the
solution would be ``extremal'' in the sense of displaying enhanced
symmetries. In this case, the maximal symmetry is just the
symmetry assumed for the ansatz, as discussed in the previous
section.

\EPSFIGURE[t]{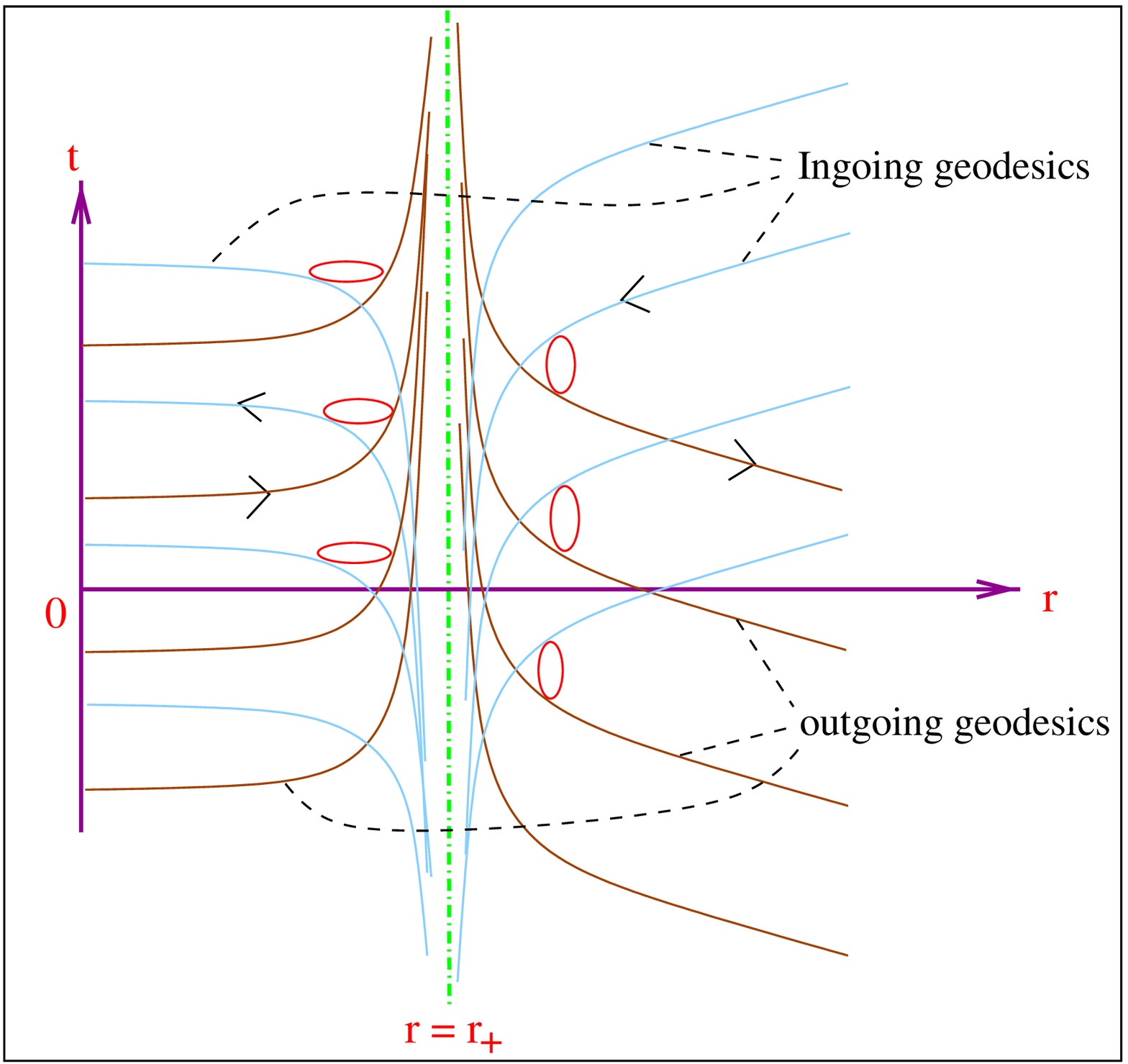, width=.6\textwidth}{Null-like
geodesics form in the simple case, $b=0$, $n=3$.\label{fig3}}

This last fact is an important difference between our solutions
and the $S$-brane solutions discussed
in~\cite{stromin,gutperle,myers}. The symmetries considered in
these works differ from those we assume, as they imposed
$\SO(n,1) \times \ISO(q+1)$ symmetry~\cite{stromin,gutperle}.
Only for the $q=0$ case, symmetries of our solution yields the
same as those discussed in~\cite{stromin,gutperle,myers}. In the
absence of the dilaton field, our solutions coincide  with the
S-branes for any $n$ (again with $q=0)$.

\paragraph{E-branes: $\eta < 0$:}

our solutions also have counterparts in the literature (in various
limits) if one makes non-standard choices for the signs of
$\alpha, \eta$ and $\lambda$. In this case, we obtain the
euclidean branes, or E-branes, discussed in~\cite{hull}. This
connection can be seen by taking ``wrong'' sign kinetic term for
the $(q+2)$-form tensor field: $\eta < 0$. With this choice, $Q$
remains real \emph{provided} we also take $(r_-r_+)^{n-1}<0$. This
leads to the low energy limit of  truncated type-II* string
theories~\cite{hull}. Taking $\eta<0$, one can take an \emph{extremal limit}: $r_-^{n-1} = -r_+^{n-1}$ in the original
solution eq.~(\ref{metric}), with a real-valued electric charge,
but in the type-II* string theories. Note that, in this limit,
the horizon  becomes singular (similar to what happens for
extremal black-branes). However, for the $E(3+1)$-brane, the
horizon is  regular and the dilaton field is well-behaved, in
close analogy with the D3-branes in type-IIB string theory.

In this case, the harmonic function $H_-$ is reduced to $1$,
rendering the space-time geometry as:
\begin{eqnarray}
ds^2 =  H_+^{(2 \slash (n-1)-b-A)} \left(- d\tau^2 +\tau^2
d\mathcal{H}^2_{n} \right)
+ H_+^{-B} \left(dr^2 + dy^2_{q} \right).
\end{eqnarray}
This solution interpolates between flat spacetime in the
asymptotic region, $\tau \rightarrow \infty$, and direct product,
$dS_{q+2} \times \mathcal{H}_n$, of a $(q+2)$-dimensional de
Sitter space and an $n-$dimensional hyperbolic space in the
near-horizon region, $\tau \rightarrow 0$. The $E(3+1)$-brane is
particularly simple, whose metric takes the form:
\begin{eqnarray}
ds^2 \vert_{E(3+1)} = -\left(\frac{r_+}{\tau}\right)^2 d\tau^2 +
\left(\frac{\tau}{r_+}\right)^2dy^2_{q+1} + r_+^2
d\mathcal{H}^2_{n} \,.
\end{eqnarray}

The above extremal solutions correspond exactly to the E-branes of
the type-II* string theories. One then expects that there ought to
be non-extremal E-brane solutions as an equivalent of the black
branes in the original type-II string theories. Some of these may
be constructed for type-II* string theories as above, by taking
both $r_-^{n-1}$ and $\eta$ negative-valued. This yields an analog
of the black-branes in type-II* string theories. In the
non-extremal case, $r_-$ is again singular. One can further find
other solutions of the type-II* string theories for $k=1$ by
taking more general values for the first-integral constants and
for the coupling parameters.

\section{Interpretation I: negative tension brane}\label{Sec4}

In this section, we shall be drawing a viable interpretation of
our solution eqs.~(\ref{metric})--(\ref{eq:F}). We will first
investigate in further detail the two conserved charges alluded in
section~\ref{Sec2}. We will see that, while the definition of the
electric charge of the source object do not pose problems, the
definition of the gravitational mass requires careful treatment.
We will then explore the spacetime geometries and causal
structures by studying geodesic motion of a test particle.

\subsection{Conserved quantities}

We start by identifying two conserved quantities as Noether
charges carried by the source branes, whose metric, dilaton field,
and $(q+2)$-form field strength are given as in
eqs.~(\ref{metric})--(\ref{eq:F}).

\paragraph{Electric charge.}

We have argued earlier that the constant $Q$, eq.~(\ref{Qdef}),
defined roughly by a flux integral of the Poincar\'e dual $n$-form
field strength ${}^* F_{q+2} = \tilde F_n$ over the
$n$-dimensional maximally symmetric space, is interpretable as a
conserved electric charge. We now elaborate the argument, and
associate the electric charge with $q$-branes located at each of
the two time-like singularities.

From the field equation eq.~(\ref{eq:F}) of the $(q+2)$-form
tensor field strength, a conserved charge density can be defined
through $d^*F_{q+2} = {}^*J$. This leads to the following
expression for the electric charge:
\begin{equation}\label{Qint}
Q=\int_{\Sigma} d\Sigma_{\mu i\dots}\nabla_{\nu}
\left( e^{-\sigma\phi} F^{\mu \nu i \dots}\right) =
\int_{\partial\Sigma} d\Sigma_{\mu\nu i\dots} e^{-\sigma\phi}
F^{\mu\nu i \dots}\,,
\end{equation}
where $\Sigma$ refers to any $(n+1)$-dimensional space-like
hyper-surface transverse to the $q$-brane. Advantage of the above
expression of the electric charge lies in the observation that the
integrand vanishes almost everywhere by virtue of the field
eq.~(\ref{eqgaf}). It does not vanish literally everywhere,
however, because the integrand behaves like a delta function at
each of the two time-like singularities. Conservation of $Q$ is
also clear in this formulation, as the second equality of
eq.~(\ref{Qint}) exhibits that $Q$ is independent of $\Sigma$ so
long as the boundary conditions on $\partial \Sigma$ are not
changed.

\looseness=-1Evaluating the flux integral eq.~(\ref{Qint}) over a space-like
hyper-surface $t=\mbox{constant}$ within either of the two static
regions (regions II and IV of figure~\ref{fig1}), we retrieve the
result eq.~(\ref{Qdef}), up to an overall normalization, for the
electric charge at each of the two time-like singularities. The
electric charge turns out \emph{equal} but \emph{opposite} for each
of the $q$-branes located at the two time-like singularities in
the fully extended space-time: $Q_\mathrm{II} = - Q_\mathrm{IV}$.
One can draw this conclusion by directly applying
eq.~(\ref{Qint}) to a choice of the space-like hyper-surface
$\Sigma$, which extends from the immediate right of the
singularity located in region II to the immediate left of the
singularity located in region IV, and passes through the
``throat'' where these regions touch (see figure~\ref{fig1}). As
this choice of the hyper-surface does not enclose the
singularities, the flux integral in eq.~(\ref{Qint}) necessarily
vanishes. This implies that the (outward-directed) electric
fluxes through the two components of the boundary,
$\partial\Sigma = \Sigma_\mathrm{II} + \Sigma_{IV}$, are equal and
opposite to one another, and so the same is true for the electric
charges which source the dilaton and the tensor fields on the two
boundaries.

We are led in this way to identify the conserved quantities, $\pm
Q$, with electric charges carried by each of the two $q$-branes
located at the time-like singularities. Which brane carries which
sign of the electric charge may be determined as follows. As
eq.~(\ref{eq:F}) defines the constant $Q$ relative to a coordinate
patch labelled by $r$ and $t$, the key observation
is that the coordinate $t$ can increase into the
future only for one of the two regions, II or IV. Then, the
charge $+Q$ applies to the brane whose static region $t$
increases into the future, and $-Q$ applies to the brane whose
$t$ increases into the past.

\paragraph{Gravitational mass.}

Recall that the metric eq.~(\ref{metric}) is static only in the
regions II and IV, but not in the regions I and III. This means
that only in the static regions is it possible to define a
conserved gravitational mass (or tension) in the usual sense for
the branes located at time-like singularities.

A procedure for evaluating the gravitational mass in
the present situation is to adopt the \emph{Komar integral}
formalism~\cite{Komar}, which cleanly associates a conserved
quantity with any Killing vector field, $\xi^\mu$, by defining a flux
integral:\footnote{We thank Gary Gibbons for interesting discussions on
this section.}
\begin{eqnarray} \label{Mdef}
K[\xi] := c\alpha \oint_{\partial
\Sigma} {dS_{\mu\nu}D^\mu \xi^\nu}\,.
\end{eqnarray}
Here, $c$ denotes a normalization constant, and $\Sigma$ is again
an $(n+1)$-dimensional space-like hyper-surface transverse to the
$q$-brane, and $\partial \Sigma$ refers to the boundary of
$\Sigma$. The Komar charge $K$ is manifestly conserved since it is
invariant under arbitrary deformations of the space-like
hyper-surface $\Sigma$ for a fixed value of the fields on the
boundary $\partial \Sigma$.

The connection between the flux integral eq.~(\ref{Mdef}) and the
more traditional representation of $K[\xi]$ as an integral over
$\Sigma$ of a current density is obtained by using the identity
$D^2 \xi^\mu = - {R^\mu}_\nu \xi^\nu$ and Gauss' law:
\begin{eqnarray}
K[\xi] = 2 c \alpha \int_{\Sigma}
{dS_{\mu}D_\nu D^\mu \xi^\nu} = \int_\Sigma {dS_\mu J^\mu(\xi)}\,;
\end{eqnarray}
where the current density
\begin{eqnarray}
J^\mu(\xi)= c \left(T^\mu_\nu\xi^\nu - \frac{1}{d-2}
T^\lambda_\lambda  \xi^\mu \right),
\end{eqnarray}
is conserved in the sense that $D_\mu J^\mu=0$. This last
expression utilizes the properties of Killing vector fields, as
well as Einstein's equations for relating $R_{\mu\nu}$ to the
total stress-tensor, $T_{\mu\nu}$. As we see explicitly later, if
$T_{\mu\nu}$ is non zero, then the value taken by $K$ can depend
on the location of the boundary $\partial\Sigma$ in
eq.~(\ref{Mdef}).

We now argue that, if we adopt the Komar integral for the
definition of the $q$-brane tension $\mathcal{T}$, the \emph{sign}
of the tension ought to be the \emph{same} for both static
regions, II and IV. This is most transparently seen for the
Schwarzschild-like solution for which $T_{\mu\nu}= 0$, by
applying the definition of eq.~(\ref{Mdef}) to the two-component
boundary of a surface, $\Sigma_t$, of constant $t$. The boundary
extends from near the singularity in region II over to near the
singularity in region IV. Then, the vanishing of $T_{\mu\nu}$
leads to the conclusion that the contribution from each boundary
component is equal and opposite: $K_\mathrm{II}(\partial_t) = -
K_\mathrm{IV}(\partial_t)$. However, since the
globally-defined time-like Killing vector is only future-directed
in one of the two regions, II or IV, local observers will
identify $\mathcal{T} = - K[\partial_t]$ in the region where
$\partial_t$ is past-directed, leading to the conclusion
$\mathcal{T}_\mathrm{II} = \mathcal{T}_\mathrm{IV}$.

To evaluate the tension $\mathcal{T} = K[\partial_t]$ in the
patch for which $\partial_t$ is future-directed, we will choose
for the hyper-surface $\Sigma$ a constant-$t$ spatial slice and
for the boundary $\partial\Sigma$ a $r =r_0$ (viz. a constant
radius) slice in the regions II and IV, respectively. It turns
out that, if $Q\ne 0$, the expression for the tension depends on
the value $r$ at which the boundary $\partial \Sigma$ is defined.
This is also true for the radius-dependent mass of the
Reissner-Nordstr\"om black-hole. Likewise, we would expect that
the gravitational mass of the $q$-brane depends on the
stress-energy of the $(q+2)$-form tensor field for which the
brane is a source if $Q\ne 0$. Explicitly, we find the tension is
given by:
\begin{eqnarray}
\frac{\mathcal{T} (r)}{V} &=& - 2\alpha (n-1)\left[ r_-^{n-1} -
k r_+^{n-1} + r_-^{n-1} (2+A-(n-1)\,b) \left(\left(\frac{r_+ }{r}
\right)^{n-1} - 1 \right) \right], \nonumber\\
&=& -\frac{(n-1)}{8\pi G}\left[ r_-^{n-1} -
k r_+^{n-1} + \frac{2 Q^2 }{(n+q)(n-1)} \left( \frac{1}{r^{n-1}} -
\frac{1}{r_+^{n-1}} \right) \right],
\label{tensione}
\end{eqnarray}
where the normalization constant $c$ has been chosen to ensure
that $\mathcal{T}/V$ takes the conventional (positive) value for
the black-brane solutions $(c=4)$, and $V$ is the volume of the
$(n+q)$-dimensional hyper-surface over which the integration is
performed. The standard normalization choices $16\pi G\eta =
1/(q+2)!$ and $16\pi G \alpha = 1$ are made in the second
equality~above.

Recall that for $k=-1,0$ in the static part of the space-time $r$
must satisfy $r \le r_{+}$, and this shows that the tension as
defined above is negative throughout the static region. For the
special case of the simple Schwarzschild-type solution discussed
in section~\ref{Sec2}, the tension becomes simply $\mathcal{T} /V
=M/V= - P/V_n$, where $V_n$ denotes the finite volume of the
$n$-sphere (so $V_2 = 4 \pi)$. As the Schwarzschild case is a
vacuum space-time $(T_{\mu\nu} = 0)$, this result is independent
of the choice of $r$, and is only non zero due to the
$\delta$-function singularity in $T_{\mu\nu}$ which the solution
displays as the time-like singularities are approached. This
again shows how the tension may be identified with $q$-branes
sitting at these singularities. For $k=1$  we recover the
standard charge dependence of the tension, in this case the
calculation is done in the region inside the second horizon.

Negative tension, $\mathcal{T}<0$, for both branes is in accord
with the form of the Penrose diagram of figure~\ref{fig1}, which,
in the static regions, II and IV, is similar to the Penrose
diagram for a negative-mass Schwarzschild
black-hole~\cite{NegMassPenrose}, or the overcharged region of
the Reissner-Nordstr\"om black-hole.  As we shall  see next,
negative-valued gravitational mass or tension is also borne out
by the behavior of the geodesics of a test particle in these
regions.

Note that the Komar integral technique used above can also be
used to compute a ``conserved'' charge in the time-dependent
regions, which may be relevant for confirming the S-brane
interpretation of the horizons of these regions. The quantity
obtained in this way involves the $ti$-components of the
stress-tensor, and defines a generalized momentum corresponding
to the symmetry under shifting $r$ in this region. We regard
however the existence of the static regions, where conserved
quantities like tension can be clearly defined, as being very
helpful in providing a physical interpretation of geometries like
S-branes.

\subsection{Repulsive geodesics}
To substantiate why the negative-tension interpretation is a
viable one, we will study geodesic motion of a test particle in
the background of the solution eqs.~(\ref{metric})--(\ref{eq:F}).
Specifically, we will be primarily interested in the $k=-1$ case,
and study geodesic motion of a massless or a massive test
particle, which couples only to the metric but not to the dilaton
or the $(q+1)$-form tensor fields. To understand the nature of the
solution beyond the static regions, we will follow the geodesic
motion of these particles starting from the past time-dependent
region III, passing through the static regions II and IV, and
eventually ending in the future time-dependent region I.

\paragraph{Null geodesics.}

In the static regions II and IV, the radial coordinate ranges
over $r<r_+$, so we consider the \emph{radial} null geodesics
defined by $ds^2 = d\mathcal{H}_n^2 = dy_q^2 = 0$. This implies
that:
\begin{equation}\label{nullmetric}
-h_-^{A+1-(n-1)b}h_+ \, \dot t^2 +
h_-^{A-1+b}h_+^{-1}\,  \dot r^2 =0\,.
\end{equation}
where the dots refer to differentiation with respect to the affine
parameter along the world-line. Thus,
\begin{equation}\label{nullgeodesic}
\frac{dt}{dr} = \frac{\dot t }{\dot r} =
\pm h_+^{-1} h_-^{-1+nb/2}\,,
\end{equation}
where the $\pm$ sign is for outgoing/ingoing geodesics.

As the regions II and IV are time-independent, a first-integral of
the geodesic equation renders energy conservation:
\begin{equation}\label{nullt}
- \xi_m {\dot x}^m = h_-^{A+1-(n-1)b}h_+ \dot t = E \,,
\end{equation}
where $\xi = \partial_t$ denotes the time-like Killing vector.
eqs.~(\ref{nullt}) and~(\ref{nullmetric}) together furnish
\begin{equation}\label{nullr}
\dot r = \pm E h_-^{-A+(n-1)b/2}\,.
\end{equation}

Though we have derived them in the static region,
eqs.~(\ref{nullr}) and~(\ref{nullgeodesic}) are applicable
equally well in other regions too. One can integrate them
numerically for a generic initial condition, and we illustrate
the result in figure~\ref{fig3}. The outgoing geodesics are those
which travel outside the horizon and pass into the region I of
figure~\ref{fig1}. Similarly, ingoing geodesics are those which
come from the past time-dependent region III in figure~\ref{fig1}.

Since $h_\pm \rightarrow \infty$ as $r \rightarrow 0$, there is
no difficulty for integrating either eq.~(\ref{nullr})
or~(\ref{nullgeodesic}) right down to $r=0$, indicating that null
geodesics reach the singularities in a finite interval of both
affine parameter and coordinate time.

\paragraph{Time-like geodesics.}

Radially directed time-like geodesic motion is characterized by
$\dot s^2 = -1$, $d\mathcal{H}_n^2 = dy_q^2 = 0$, and so:
\begin{equation}\label{timemetric}
-h_-^{A+1-(n-1)b}h_+ \dot t^2 +
h_-^{A-1+b}h_+^{-1} \dot r^2 = -1 \,.
\end{equation}
Combining this with the first-integral of energy conservation,
eq.~(\ref{nullt}), we find the following condition for time-like
geodesics in all regions:
\begin{equation}\label{timer}
\dot r = \pm \left( E^2 h_-^{-2A+(n-2)b} -
h_-^{-A- b+1}h_+ \right)^{1/2},
\end{equation}
where again the sign is $+$ for outgoing and $-$ for incoming radial
time-like geodesics.

As before, the geodesic equations can be integrated numerically in
the general case, but it is clear that the observer takes a finite
proper-time to reach the horizon $(h_+ \rightarrow 0)$, across
which the observer can pass freely. In terms of the coordinate
time, we have:
\begin{equation}\label{timegeodesic}
\frac{dt}{dr} = \pm \frac{E  h_-^{(n-1)b -A}}
{h_+h_-\sqrt{E^2  h_-^{-2A+(n-2)b} - h_-^{1-A-b}h_+}}\,.
\end{equation}
The integral of eq.~(\ref{timegeodesic}) diverges as $h_+
\rightarrow 0$, so we see that it takes an infinite time for a
particle to reach the horizon as seen by a static observer inside
the horizon $(r < r_+)$.

\EPSFIGURE[t]{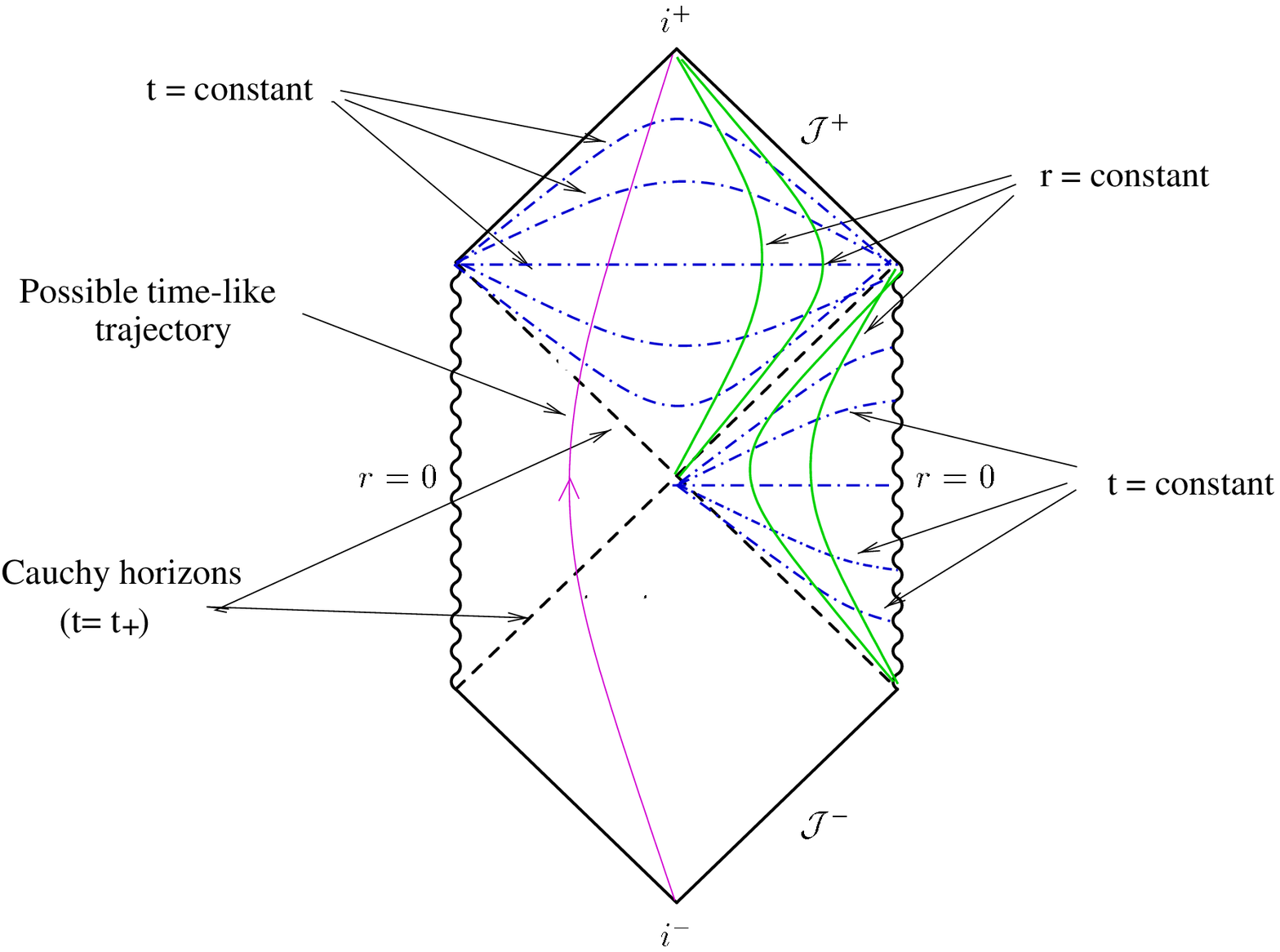, width=.7\textwidth}{Typical
time-like trajectory. Shown also are constant-$r$, constant-$t$
surfaces and the Cauchy horizon of the solution.\label{penrose1}}

On the other hand, as $r \rightarrow 0$, $h_\pm \sim 1/r^{n-1} \rightarrow
\infty$. In this limit, the first term inside the square root of
eq.~(\ref{timer}) grows slower than the second term, which renders
the square-root to become complex-valued if $r$ becomes
sufficiently small. We see from this that an infalling time-like
geodesic never hits the singularity. Instead the infalling
observer reaches a point of closest approach, $r_m>0$, at which
the square root in eq.~(\ref{timer}) becomes zero, and reflected
outward. The turning point for a time-like geodesic is given by
the value $r_{c}$ of the coordinate $r$, for which the following
expression holds:
\begin{eqnarray}
r_{c}=\frac{r_{+}}{(1+E^{2}[h_{-}(r_{c})]^{(n-1)b+A-1})^{1
\slash (n-1)}}\,.
\end{eqnarray}
Note that $r_{c}$ is always between $0 \le r_c \le r_{+}$. For
example, for $r_{-}=0$, and $k=-1$, we have that
\begin{eqnarray}
r_{c}=\frac{r_{+}}{(1+E^{2})^{1 \slash (n-1)}}\,.
\end{eqnarray}

\paragraph{Gravitational repulsion.}

We see from the above considerations concerning geodesic motion of
a test particle that the two time-like singularities act
gravitationally as \emph{repulsive} centers, as no infalling
time-like geodesic can hit them.\footnote{However, infalling null
geodesics can hit the singularity.} In this  sense, the space is
time-like (although not null) geodesically complete. Observers
who originate in the remote past --- region III --- enter one of
the static regions by passing through the past horizon, and then
leave this through the future horizon of the future
time-dependent region, I. This resembles what happens in other
geometries, such as the Reissner-Nordstr\"om black-hole.

Putting together the above results, we are led to draw the
following conclusion. Static observers in regions II and IV are
those whose world trajectories follow lines of constant $r$, and
these observers have proper accelerations which are directed
towards the nearest singularity. We find the following expression
for the proper acceleration, for $k=-1,0$ and $q=0$. We have, in
the coordinates adopted,
\begin{eqnarray}
a^{r}=-\frac{(n-1)h_{+}h_{-}^{1-b}}{2 r^{n}}
\left[\frac{r_{+}^{n-1}}{h_{+}}+\frac{r_{-}^{n-1}}{h_{-}}\right],
\end{eqnarray}
so the value of the acceleration is always negative.

The singularities themselves are special instances of these
observers for whom $ r \rightarrow 0$, in which limit the proper
acceleration becomes infinitely large. As discussed in the
Introduction, this behavior does not contradict with the
equations of motion for the branes at the singularities since for
negative-tension branes these do \emph{not} imply motion along a
geodesic (or otherwise) within the space-time.

This is in contrast to what is found for accelerating
positive-mass particles, as described by the C-metric. For this
metric, the particle world-lines are also not geodesics, so the
particles follow trajectories which are not self-consistently
determined by the fields which the particles source. For
positive-mass particles this inconsistency shows up through the
appearance of nodal defects, which are conical singularities along
the line connecting the two particles. These singularities are
interpreted as being the gravitational influence of whatever
additional stress-energy is responsible for the particle
motion~\cite{CMetricD,CMetricInt}.

\section{Interpretation II: time-like wormhole}\label{scbounce}

Comparison with the Schwarzschild black-hole permits another
interpretation of our solution. After re-expressing our solutions
eqs.~(\ref{metric})--(\ref{eq:F}) in conformal frame, the geometry
of the $n$-dimensional slices turns out that of a \emph{time-like
bounce}. On the other hand, the  scale factor for the $r$
coordinate resembles an object localized in time, and so is a kind
of \emph{time-like kink}. Such bounce/kink behavior would help
explain what precisely the S-brane configuration is.

We shall be interested in foliating the geometry with respect to
the time in the time-dependent regions, I and III. We will be
finding that the geometry exhibits bounce and kink behavior for
the symmetric space and the radial direction, respectively.

\subsection{Einstein-Rosen wormhole: a review}
We begin by recapitulating the interior dynamics of the
Schwarzschild black-hole relevant for our foregoing discussions.

Consider the maximally extended space-time of the Schwarzschild
black-hole. We are interested in describing time-evolution of the
space-time geometry. We may foliate the space-time as a stack of
constant $t$ surfaces. Then, the space-time at sufficiently early
epoch consists of two disconnected asymptotically flat components,
each containing a space-like singularity surrounded by a past
horizon. The two components evolve and, at some early epoch, the
two singularities join together and smooth out by forming a
``wormhole'' connecting the two components. The wormhole neck
widens, reaching a maximal proper size $r = 2 M$ at the
time-symmetric point $t = 0$. This is the instance when the
wormhole neck is instantaneously static and the event horizon of
the two components join instantaneously. Evolving further, the
wormhole neck recontracts, eventually pinching off as the two
singularities reappear and the space-time disconnects.

Two remarks are in order. First, as it is evident from the Kruskal
coordinates, the process of wormhole formation and recollapse
occurs so rapidly that it is impossible to traverse the wormhole
and communicate between the two asymptotic regions without
encountering the singularity. Second, the picture of the
time-evolution depends on the foliation. Consider, for instance,
an alternative foliation illustrated in the right of
figure~\ref{wormhole}. In this case, the geometry starts as a
space-like singularity in the asymptotic past, grows out as a
hyperboloid, reaches a \emph{maximal} neck size of the hyperboloid,
and recollapses to a space-like singularity in the asymptotic
future. See figure~\ref{wormholeview} for the comparison.

\EPSFIGURE[t]{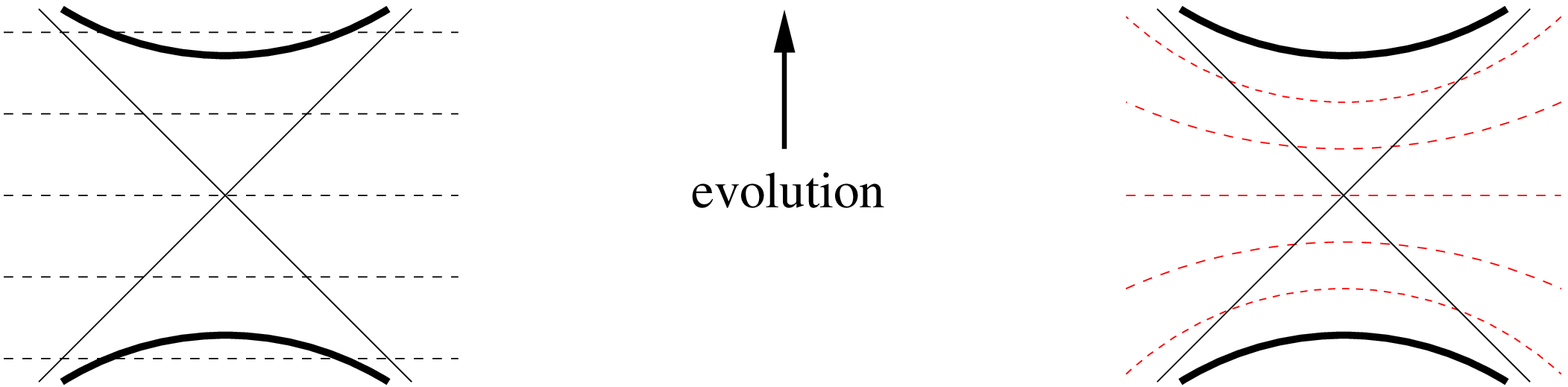, width=.9\textwidth}{Two possible
foliation of the maximally extended space-time of the
Schwarzschild black-hole. Both cases lead to the Einstein-Rosen
bridge connecting the two static regions. \label{wormhole}}

\EPSFIGURE[t]{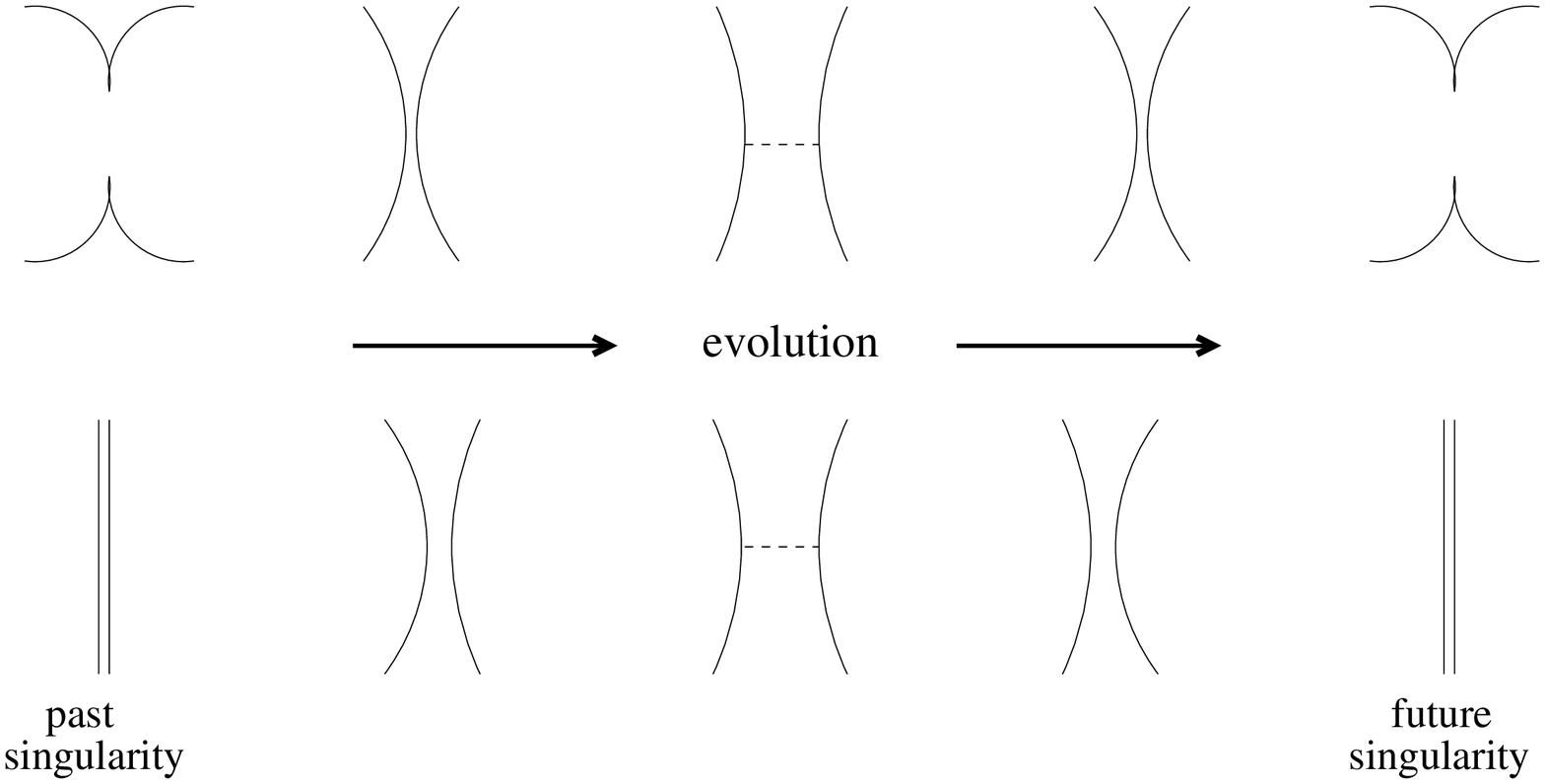, width=.9\textwidth}{Cartoon view
of time evolution of the Einstein-Rosen bridges. The upper/lower
sequence corresponds to the evolution for the left/right choice of
the foliation in figure~\ref{wormhole}. \label{wormholeview}}

A natural question is whether a foliation similar to the
Schwarzschild black-hole is possible for our solution as well. We
find that it is, although a marked difference would be that the
time-evolution is with respect to the regions outside the horizon
(inside out compared to the Schwarzschild black-hole case) and
details of the evolution are somewhat different for $k=-1$ and
$k=0$ branes, although the properties we end up finding turn out
similar. As such, we will again separate the discussion for the
$k=0$ and $k=-1$ cases.

\subsection{The $k=-1$ brane}
Consider, for simplicity, the case $r_{-}=0$ and $q=0$, for which
the singularities are point-like.\footnote{The case $r_-\neq 0$
for $b=0$ can also be integrated analytically, but gives rise to a
more complicated result.} Recall that the metric in the original
coordinates is
\begin{eqnarray}
d s^{2} = - \frac{1 }{h_{+}} d t^{2}+ h_{+}\, d r^{2} +
t^{2} d x_{n,-1}^2\,,
\label{simplif}
\end{eqnarray}
where $h_{+}=1-({r_{+}}/{t})^{n-1} $. We now rewrite this metric
in terms of the conformal time $\eta$ (not to be confused with
the normalization constant used in earlier sections) as
\begin{eqnarray}
d s^{2} = C^2(\eta)\left[- d\eta^2 + d x_{n,-1}^2\right]
+ D^2(\eta)dr^2\,,
\end{eqnarray}
where the conformal time is defined by
\begin{eqnarray}
C(\eta)= t(\eta) =
r_+ \cosh ^{2/(n-1)}\left[\frac{(n-1)}{2} \eta \right] \ge r_+\,,
\end{eqnarray}
and so $\eta$ ranges over $ -\infty < \eta < \infty $. Then, the
scale factor for the $r$-direction becomes
\begin{eqnarray}
D(\eta) = \tanh \left[\frac{(n-1)}{2}\eta \right]
\end{eqnarray}
and has the same functional dependence for all values of $n$.
These expressions exhibit the bouncing structure of the
$(n+1)$-dimensional space and the (time-like) kink structure of
the radial dimension. We illustrate the  behavior of the scale
factor in figure~\ref{figurekminus} for the example of $n=3$ and
$r_+=1$.

\FIGURE[t]{\epsfig{file=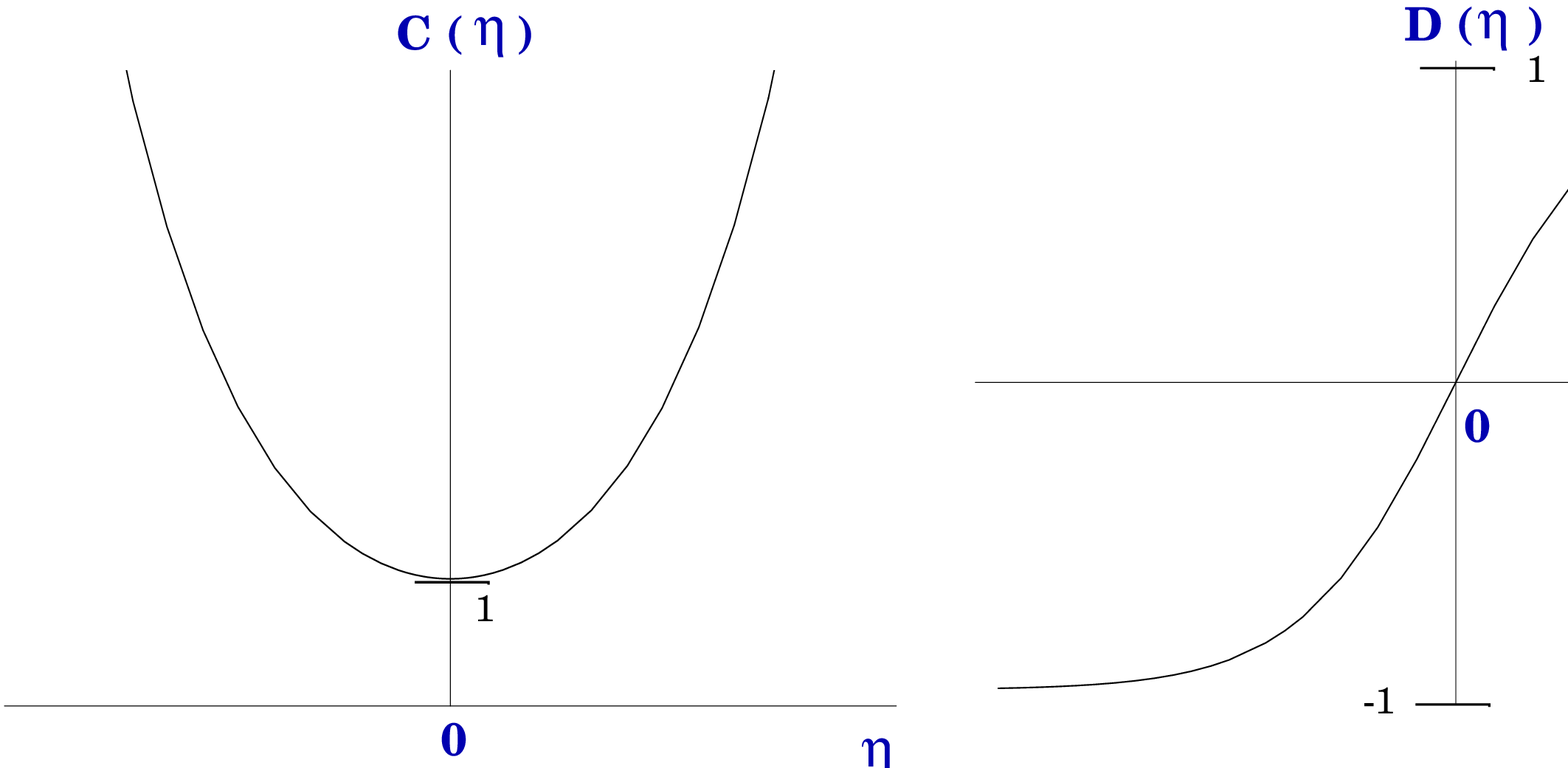, width=5in}%
\caption{A bounce and kink for the $k=-1$ brane.}\label{figurekminus}}

\subsection{The $k=0$ brane}
In this case, we cannot take a vanishing charge $(r_-=0)$, as then
$h_-$ would vanish too. We instead concentrate on the limit $b=0$
and $q=0$ but $r_- \neq 0$.  The starting metric has the same form
as eq.~(\ref{simplif}) and $h_+$ has the same form also, but now
$h_{-}= (r_-/t)^{n-1}$. In terms of the conformal time $\eta$, the
metric becomes, as before,
\begin{eqnarray}
d s^{2} = C^2(\eta)\left[- d\eta^2 + d x_{n,0}^2\right] +
D^2(\eta)dr^2\,,
\end{eqnarray}
but now with the conformal time defined by
\begin{eqnarray}
C(\eta)= t(\eta) = r_+ \left[1 + \frac{(n-1)^{2} r_-^{n-1}}{4
r_+^{n-1}}  \eta^{2} \right]^{\frac{1}{n-1}} \ge r_+\,,
\end{eqnarray}
and ranging again over $- \infty < \eta < \infty $. The scale
factor for $r$ is similarly obtained for general $n$, and is
\begin{eqnarray}
D(\eta) = \frac{2(n-1) r_-^{(n-1)} \eta }{4r_+^{n-1}
+ (n-1)^2 r_-^{(n-1)}\,  \eta^2}\,.
\end{eqnarray}
We see once again the bounce behavior of the $(n+1)$-dimensional
symmetric space and the kink behavior of the scale factor for the
$r$-direction. We can see this clearly in figure~\ref{figurek0},
where as before we plot an example for $n=3$. Note that,  at
$\eta=0$ viz. at $t = r_+$, nothing special happens to the
$(n+1)$-dimensional subspace, but the scale-factor for the
``extra dimension'', $r$, degenerates to zero!

\FIGURE[t]{\epsfig{file=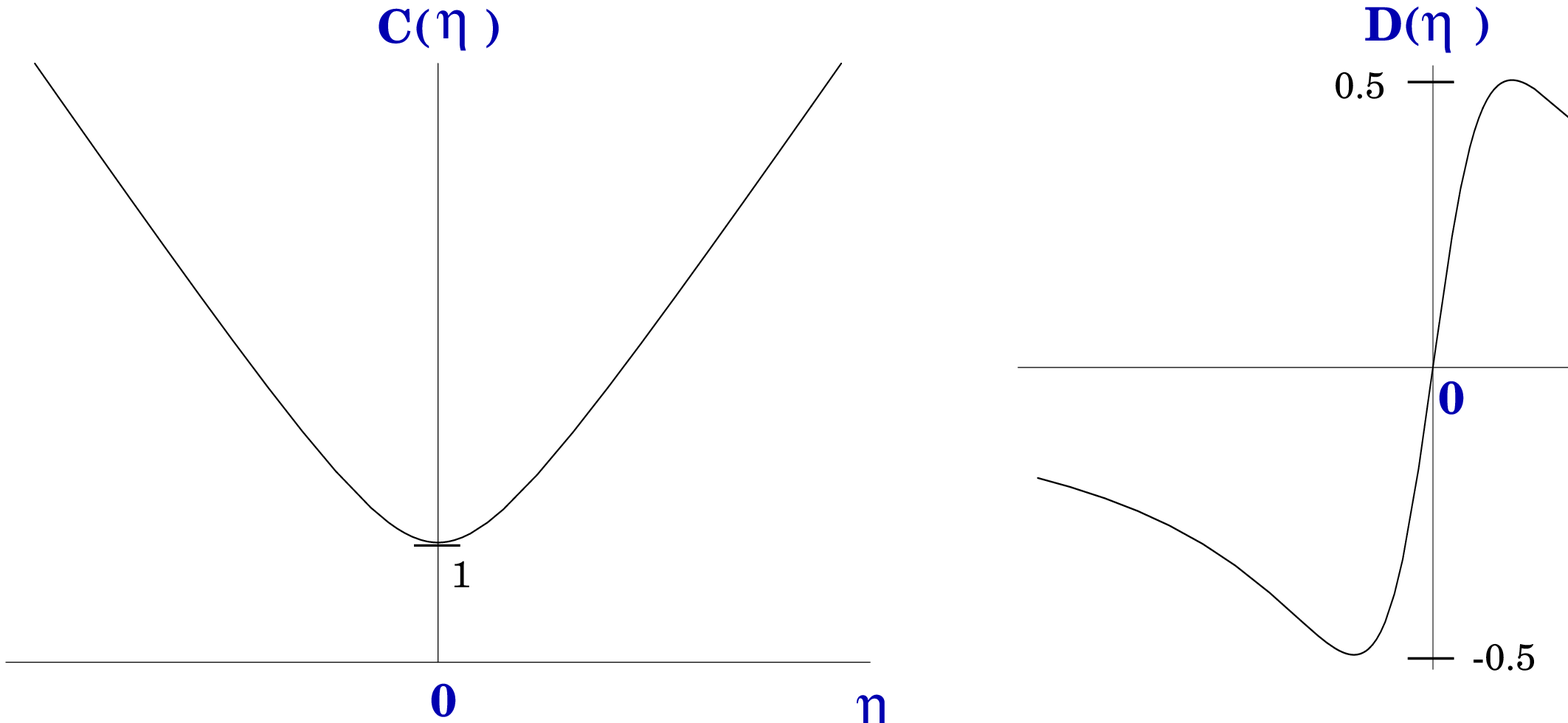, width=5in}%
\caption{A bounce and kink for the $k=0$ brane.}\label{figurek0}}

\subsection{Cosmological bounce/kink and time-like wormhole}
As anticipated, the cosmological bounce behavior of our solution
offers yet another physical interpretation: our solution is
reminiscent of a \emph{time-like} version of the Schwarzschild
wormhole or Einstein-Rosen bridge, which connects the two
asymptotically flat regions in the maximally extended Kruskal
coordinate space-time. Our solution corresponds to
a~$\pi/2$-rotation of the foliation illustrated in
figure~\ref{wormhole} in the sense that the two time-dependent
regions  --- instead of the static regions --- are connected by a
\emph{time-like wormhole}. Note that, according to
figures~\ref{figurekminus} and~\ref{figurek0}, the geometry of
each fixed $r$ slice start out contracting, reaching the minimum
volume, and subsequently expanding.

The bounce/kink interpretation of our solution fits also nicely
with the interpretation that, in particular cases, our solution
reduces to the S-brane (as alluded earlier), and with the proposal
that the S-branes are time-like kinks. Our solution clarifies the
proposal further in that the S-brane is in fact located at the
horizon $r_+$.

\subsection{Comparison with Reissner-Nordstr\"om black-hole}
An attentive reader would have not missed the resemblance between
our solutions and (part of) the spacetime of the
Reissner-Nordstrom black-hole. More specifically, if we let the
outer horizon of the Reissner-Nordstrom black-hole go to
infinity, then the geometry and the Penrose diagram of the two
space-times are the same.

We believe that the previous interpretation of the Kruskal diagram
for our $k=0, -1$ solutions in terms of interactions due to
negative-tension objects remains valid also for the non-extremal
Reissner-Norsdr\"om black-hole in four dimensions, whose horizon
is given by $S_2$ and $k=1$ (switching off the dilaton,
see~\cite[pg.~158]{hawking}). In our analysis, the time-dependent
region, the region between the inner and the outer horizons, is
interpretable as a destabilization of the space-time due to the
combined gravitational field of two negative-mass objects.
Inspecting the Penrose diagram of the non-extremal
Reissner-Norsdr\"om black-hole, one notes that the same
considerations are applicable. First of all, the two singularities
in the region inside the inner horizon, where the space-time is
static, still exhibit \emph{opposite charges} and \emph{equal} but
\emph{negative masses}. The negative value of the mass obtained
from the Komar integral calculation is essentially due to
contributions coming from the electromagnetic field.

The past light-cone of an observer in the ``time-dependent'' region
--- the region between the inner and the outer horizons --- is aware
of both the negative-mass objects located inside the inner
horizons: the simultaneous repulsion of the two objects propel the
observer toward increasing values of the coordinate $r$. Once the
observer crosses the horizon corresponding to $r=r_{+}$, entering
into the region outside the outer horizon, the observer's past
light-cone does not see any longer two negative-mass objects, but
only one. The interaction with only one object is not sufficient
to destabilize the space-time. In the asymptotically flat static
region outside the outer horizon, the Komar integral calculation
gives a positive mass object: indeed, the effect of the
electromagnetic field is suppressed in comparison with the
gravitational one.

Passing to a conformal frame, constructing the wormhole solution
connecting the time-dependent regions of the metric, one finds a
``bounce structure'' with a periodic cosine dependence, instead of
the hyperbolic-cosine  one obtained for $k=-1,0$, describing in
this way a cyclic universe (for related ideas see for
instance~\cite{cyclic}).

\section{Stability, singularity and thermodynamics}\label{Sec5}

An immediate question is whether our solution
eqs.~(\ref{metric})--(\ref{eq:F}) is stable. In this section, for
definiteness, we shall be taking again the particular solution:
$k=-1$ brane with $q=0$, and make a first step toward the
complete stability analysis, both at classical and quantum
levels. At the same time, based on these results, we draw
definitive statements concerning the physical nature of the
time-like singularities inherent to our solution.

\subsection{The Cauchy horizon}
An analysis of the stability of --- or the particle production by
--- a given space-time starts with initially-small fluctuations of
the fields involved, and propagates them forward in time
throughout the space-time. The set-up therefore presupposes that
the initial-value problem is well-posed. In the space-time of
eq.~(\ref{metric}), this is not clear as there exists a \emph{Cauchy horizon}, which separates the past time-dependent region
III from the static regions II and IV. The Cauchy horizon exists
because initial conditions specified in region III do not
uniquely determine the future evolution of the fluctuation
fields. They do not do so because all points after the Cauchy
horizon have at least one singularity in their past light cone,
and so can potentially receive signals from these singularities.
This implies that a unique time evolution of a field fluctuation
from the past time-dependent region III into the future
time-dependent region I must also involve a specification of some
sort of boundary condition at the location of the two time-like
singularities.

From the perspective of brane physics, the existence of such
Cauchy horizons is physically reasonable. Imagine that the
time-like singularities are the positions of real branes. There
then exists a possibility that these branes might emit radiation
into the future time-dependent region I, and the possible choices
for boundary conditions at the singularities simply encode the
possible emission processes which can occur on branes'
world-volume. A well-posed time-evolution problem in the embedding
space-time thus requires specification as to whether or not the
branes are emitting or absorbing radiation.

When necessary, we shall choose the simplest possible brane
boundary condition: we assume the brane neither emits nor absorbs
any radiation.

\subsection{The Klein-Gordon equation}
We first consider the Klein-Gordon equation for a scalar field
propagating in the background eqs.~(\ref{metric})--(\ref{eq:F}),
with particular attention paid to these equations' limiting
behavior at asymptotic infinity, and near the horizons. We then
explore some relevant properties of the solutions in these
regions.

Consider the Klein-Gordon equation of a massive scalar field:
\begin{eqnarray*}
- \frac{1}{\sqrt{g}}\partial_{M}\left[\sqrt{g}g^{MN}\partial_{N}\right]
\psi + M^{2}\psi = 0
\end{eqnarray*}
in the time-dependent regions I and III. Adopting the isotropic
coordinates, the equation is given by
\begin{eqnarray}
- \frac{1}{\sqrt{g}}\partial_{\tau} \left[ \sqrt{g}
g^{\tau\tau} \partial_{\tau} \right] \psi -  g^{rr}
\partial_{r}^2 \psi -  \frac{1}{\omega^2 \sqrt{h}} \partial_{i}
\left[\sqrt{h} h^{ij} \partial_{j} \right] \psi + M^{2}\psi = 0\,.
\end{eqnarray}
Here, for clarity, we denote $h_{ij}(x)$ for the metric on the
$n$-dimensional maximally-symmetric hyperbolic space
$\mathcal{H}_n$, whose coordinates are $x^i$, and write
$g_{ij}(\tau,x) = \omega^2(\tau) h_{ij}(x)$. The relevant metric
components are:
\begin{eqnarray}
g_{\tau\tau} &=& - \left(\frac{ H_-}{H_+} \right)^{b}
\frac{H_+^{2/(n-1)} }{H_-} \,, \nonumber \\
g_{rr} &=& \left(\frac{ H_-}{H_+} \right)^{1-(n-1)b}
\frac{1}{H_+} \,,\nonumber\\
\omega^2 &=& \tau^2 \, \left(\frac{ H_-}{H_+} \right)^{b}
H_+^{2/(n-1)} \,.
\end{eqnarray}
The functional form of the metric involved permits separation of
variables, so we take $\psi(r,t,x) = \mathrm{e}^{i P\,r} \, f(t)
\, L_K(x)$, where $P$ and $K$ are separation constants determined
by the eigenvalue equations:
\begin{eqnarray*}
- \partial_r^2 e^{iP\, r} = P^2 e^{iP\, r}\,, \qquad \mbox{and}
\qquad - \frac{1}{\sqrt{h}} \partial_{i} \left[ \sqrt{h} h^{ij}
\partial_{j} \right] L_K = K^2 L_K\,.
\end{eqnarray*}
Both eigenvalue equations can be solved explicitly, and
delta-function or $\mathcal{L}_2$ normalizability of the solutions
require both $P^2 \ge 0$ and $K^2 \ge 0$. The temporal eigenvalue
equation then becomes:
\begin{equation}\label{KGteq}
- \frac{1}{\sqrt{g}} \frac{d}{d\tau}
\left[\sqrt{g} g^{\tau\tau} \frac{df }{d\tau} \right] +
\left[g^{rr} P^2  + \frac{K^2}{\omega^2}  + M^{2}\right] f = 0\,.
\end{equation}

\paragraph{Asymptotic past/future.}

In the asymptotic future and past regions I and III, $\tau \rightarrow
\infty$, so the metric becomes flat with $H_\pm \rightarrow 1$, and the
mode functions go over to standard forms. In this limit,
eq.~(\ref{KGteq}) is reduced to
\begin{equation}\label{KGt}
\ddot f + \frac{n }{\tau} \dot f + \left( P^2 + M^2 +
\frac{K^2}{\tau^2} \right)  f = 0\,,
\end{equation}
where the dots represent  derivatives with respect to $\tau$. The
solution is expressible in terms of the Bessel functions:
\begin{equation}
f(\tau) = \tau^{(1-n)/2}\left[\alpha_1  J_{y}(\rho\,\tau) +
\alpha_2  Y_{y}(\rho\,\tau) \right],
\end{equation}
where $y = -\frac12 \sqrt{(n-1)^2-4\,K^2}$, the $\alpha_1,
\alpha_2$ are constants of integration, and the parameter in the
argument is $\rho = \sqrt{P^2 + M^2}$.

At future infinity in the time-dependent region I (or past
infinity in region III), we find the asymptotic behavior of the
solution is $f(\tau) \sim \tau^{-n/2} e^{\pm i P\tau}$, if $P \ne
0$. If $P=0$ then $f(\tau) \sim \tau^{a_\infty}$, with
\begin{equation}
a_\infty = - \frac12 \left[(n-1) \pm \sqrt{(n-1)^2 - 4 K^2}
\right].
\end{equation}
These solutions are oscillatory for all $K^2 > \frac14 (n-1)^2$,
and do not grow with $\tau$ for large $\tau$ so long as $K^2 \ge
0$.

\paragraph{Near-horizon limit.\protect\footnote{In this subsection,
we relax the restriction to $k=-1$, and treat all possible cases
on equal footing.}}

Near the horizon, $\tau \rightarrow 0$ and the asymptotic form is
governed by the limits $H_+ \rightarrow (r_+/\tau)^{n-1}$ and
$H_- \rightarrow (\overline{r} / \tau)^{n-1}$, with
$\overline{r}^{n-1} = ( r_-^{n-1} - k r_+^{n-1})$. The metric
functions therefore reduce to $g_{\tau\tau}
\rightarrow \alpha_\tau \tau^{n-3}$, $g_{rr} \rightarrow \alpha_r
\tau^{n-1}$ and $\omega \rightarrow \alpha_\omega$. The precise
values of the constants $\alpha_\tau, \alpha_r$ and
$\alpha_\omega$ are not required, apart from the following ratio:
\begin{eqnarray}
\frac{\alpha_\tau}{\alpha_r} = r_+^2 \left(\frac{
\overline{r} }{r_+} \right)^{(nb-2)(n-1)} = r_+^2
\left[\left(\frac{r_- }{r_+} \right)^{n-1} - k \right]^{nb-2}.
\end{eqnarray}
With these limits, the Klein-Gordon equation becomes, in the
near-horizon limit:
\begin{equation}\label{KGthor}
\ddot f + \frac{1}{\tau} \dot f + \left[\frac{\alpha_\tau P^2}{\alpha_r }
\frac{1}{\tau^2}+ \alpha_\tau \tau^{n-3}
\left( M^2 + \frac{K^2}{\alpha_\omega^2}
\right) \right] f = 0\,,
\end{equation}
If $P\ne 0$, then the solutions are oscillatory, having the form
$f \sim \tau^{a_0}$, with $a_0 \!=\! \pm i P
\sqrt{\alpha_\tau/\alpha_r}$. If $P=0$, then a similar argument
shows that the solutions are non singular as $\tau \rightarrow 0$.

The logarithmic singularity which is implied by the form
$\tau^{a_0}$ found above has a familiar source, which is most
easily seen by transforming to ``tortoise'' coordinates: $t_* = t
+ r_+ \log[(t/r_+) -1]$, whose range is $-\infty < t_* < \infty$,
with $t_* \rightarrow -\infty$ corresponding to the horizon due
to the logarithmic singularity as $t \rightarrow r_+$. In terms
of the tortoise coordinate, the dominant part of the Klein-Gordon
equation governing the $r$ and $t_*$ dependence of $\psi$ becomes
\begin{eqnarray*}
\left(- \partial_{t_*}^2 +\partial_r^2 \right) \psi = 0\,.
\end{eqnarray*}
This simple wave equation describes waves propagating in \emph{both} directions across the horizon. Note that the mass term drops
out of these asymptotic expressions, and so, near the horizon, a
massive field behaves like a massless one, approximately
propagating along the light-cone. Just as for our discussion of
the geodesics, these ingoing and outgoing modes describe motion
into and out of the static regions, II and IV, evolving from the
past time-dependent region III and to the future time-dependent
region I.

\subsection{Classical stability}\label{SecCS}
We may now ask whether our solutions
eqs.~(\ref{metric})--(\ref{eq:F}) are classically stable in the
time-dependent regions, I and III. Classical instability is
understood here to mean that initially-small fluctuations grow
much more strongly with time than does the background metric.
Although a complete stability analysis is beyond the scope of
this paper, we perform the first steps here for scalar
fluctuations which are governed by the Klein-Gordon equation. For
simplicity, we focus in this discussion on the massless case,
$M=0$.

There are two parts to be studied for the stability analysis.
First, identify the modes which grow uncontrollably, and then
determine whether well-behaved initial conditions can generate the
uncontrollably growing modes, if these exist. In the present
instance, we have just seen that the asymptotic forms for the
Klein-Gordon solutions do not include any growing modes, due to
the conditions $P^2 \ge 0$ and $K^2 \ge 0$, which follow from the
normalizability of the spatial mode functions.

Potentially more dangerous are growing metric modes near the past
horizons, which divide the past time-dependent region III from the
static regions II and IV. These are more dangerous because of the
infinite blue-shift which infalling modes from the region III
would experience as they fall into the horizon. This blue-shift
boosts their energy (as seen by infalling observers) to
arbitrarily large values, and one suspects that such large energy
densities drive runaway behavior in the gravitational modes, much
as has been found to be so for the \emph{inner horizon} of the
Reissner-Nordstr\"om black-hole. Naively we might have expected
that the horizon in our case could be better behaved than the
Reissner-Nordstr\"om case~\cite{hawking, chandra} due to the
presence in that case of the asymptotically flat static region
from where the signals sent to the horizon are infinitely
blue-shifted. However in our case that region is absent.
Nevertheless this  does not guarantee the stability of the horizon
and a careful stability analysis needs to  be performed.

As a preliminary estimate of whether such an instability does
exist, we compute the energy, $E = - u^m \partial_m \psi$ of the
Klein-Gordon modes considered above as seen by an observer whose
velocity, $u = M \partial_t + N \partial_r$, is well-behaved as it
crosses the horizon. The normalization condition $u^2 = -1$ in the
vicinity of the horizon allows a determination of how $M$ and $N$
must behave as $\tau \rightarrow 0$ (in isotropic coordinates) in
order to remain non singular. We find in this way $u^2 \sim
-\alpha_\tau M^2 \tau^{n-3} + \alpha_r N^2 \tau^{n-1}$, which is
regular near $\tau \rightarrow 0$ provided $M \sim
\tau^{-(n-3)/2}$ and $N \sim \tau^{-(n-1)/2}$ near the horizon.
With this choice, one then finds
\begin{eqnarray}
-E = M \partial_\tau \psi + N \partial_r \psi \sim \psi
\tau^{ - (n-1)/2} \,.
\end{eqnarray}
Using the asymptotic solution found below eq.~(\ref{KGthor}): $\psi
\sim \tau^{a_0}$ with $a_0 = \pm i P \sqrt{\alpha_\tau/\alpha_r}$,
we see that $E \rightarrow \infty$ as the horizon is approached.
This suggests that the stress-energy density of the mode under
consideration diverges as well in this limit. As such, this mode
is likely to destabilize the metric modes near the past horizon,
much like what is found for the Reissner-Nordstr\"om black-hole
near $r = r_-$. Notice that if the horizon were stable, we would
have a counter-example to the strong version of the cosmic
censorship hypothesis, since observers coming from the past
cosmological region III could examine the singularity without
having to fall into it (see for instance~\cite{robetal}).

There is a second kind of instability of the Reissner-Nordstr\"om
black-hole, which our solutions do \emph{not} share. This second
stability problem for the Reissner-Nordstr\"om horizon is seen as
soon as the Einstein-Maxwell system is extended to include also a
scalar field, e.g.\ Einstein-dilaton-Maxwell system: in this
case, the inner horizon turns into a genuine singularity. A
similar problem does not arise for our solution, since our
solution is already a solution to the combined
Einstein-dilaton-$(q+2)$-form Maxwell system. We can see
explicitly that turning the dilaton on or off does not change the
structure of the horizon. Of course, a more detailed calculation
of the metric modes is required to establish definitively whether
this instability does really arise.

We see in this way that the horizons to the past of the static
regions are likely to be unstable to becoming singularities in
response to small perturbations. On the other hand, we do not
expect a similar instability for the horizons to the future of the
static regions. Certainly, a more detailed stability analysis of
these space-times is desirable.

\subsection{Issue of quantum stability}
Before proceeding describing some aspects of particle production
on these spacetimes, we first pause to remind the reader of some
general stability issues.

\paragraph{The Hawking-Ellis vacuum stability theorem.}

Hawking and Ellis~\cite{hawking} have proposed a generalization
to curved space of the familiar flat-space stability condition
that a system's energy must be bounded from below. They propose
that the energy of a physically sensible theory should be
required to satisfy the following positivity conditions, at least
on classical macroscopically averaged scales. Specifically, for
an arbitrary future-directed, time-like unit-vector, $t^\mu$, the
corresponding energy flux vector $E^\mu = - T^{\mu \nu} t_\nu$
ought to be null- or time-like and future-directed:
\begin{eqnarray}
\vert E^\mu \vert^2 \le 0\,, \qquad
\mbox{and} \qquad - E^\mu t_\mu = T^{\mu \nu} t_\mu t_\nu \ge 0\,.
\end{eqnarray}
This last inequality implies that the energy density seen by all
observers is non negative.

Physically, this condition ensures the vacuum is stable against
the spontaneous pair creation of positive- and negative-mass
objects. Given that our solution is interpreted here in terms of
objects --- more precisely, a pair of equal-tension $q$-branes ---
whose tensions clearly violate the weak energy condition, one
might be concerned about instability due to runaway particle
production.

It is important in this kind of discussion to distinguish
carefully between the energy density defined by the stress tensor
of the fields of the problem, and the tension of the $q$-branes
which are their sources. For field fluctuations it is the local
field stress energy which is important, and although the
$q$-brane tension is negative, the field stress energy is
everywhere positive or zero. For instance, the simple,
four-dimensional Schwarzschild-type solution studied in
section~\ref{Sec2} has a vanishing energy-momentum tensor except
at the location $r=0$. The geodesically complete spacetime of the
solution, however, does not include this point, implying that the
energy condition is satisfied \emph{globally}.

Further insight is provided by the consideration of the
non-extremal Reissner-Nord\-str\"om black-hole in four dimensions,
the situation elucidated in section 4.3. There, we have shown
that the region inside the outer horizon exhibits precisely the
same physical characteristics  as our solutions: the region
between the outer and the inner horizon is cosmological, while
the region inside the inner horizon corresponds to the static
region, and the black-hole singularity inside the static region
is \emph{time-like}. We have argued that the Komar mass is
negative if measured inside the static region, i.e. inside the
inner horizon. The negative-mass, however, does not imply
violation of the energy condition. This is because, as is
well-known, the stress-tensor of the electromagnetic field is
well-behaved everywhere, and can be related to the the local mass
$M(r)$ via, for example, $dM(r) \slash d r = 4 \pi r^2 T_{tt}$.
Thus, though $T_{tt}$ is positive everywhere, the local mass
$M(r)$ can become negative inside the inner horizon because the
large electromagnetic field digs up a deep gravitational
potential well. The latter is precisely what renders the Komar
mass negative when measured inside the inner horizon. By the same
line of reasoning, one can understand why the Komar mass turns
out positive \emph{if} measured outside the outer horizon.

Indeed, we have a situation similar to the above cases: the
stress-tensor of matter fields in the right-hand side of
eq.~(\ref{einstein}) are well-defined, and are positive-definite.
Despite being so, the Komar mass, defining a local mass, can
become negative inside the horizon, as the positivity of the
matter stress-tensor imposes the positivity of radial variation of
the tension but not that of the tension itself.

\paragraph{Absence of pair production.}

The above arguments based on the Schwarzschild-type solution and
the non-extremal Reissner-Nordstr\"om black-hole interior indicate
that the energy condition and hence the vacuum stability condition
are satisfied by the more general solution,
eqs.~(\ref{metric})--(\ref{eq:F}). Nevertheless, it would be
highly desirable to find yet another argument for the vacuum
stability. We believe that the existence of the orientifolds  in
string theory point to a possible resolution to the question. The
orientifolds, being a class of objects carrying negative tension,
are vulnerable to the vacuum stability theorem as well, yet they
are perfectly well-behaved structures in string theory. A good
example is the orientifold 6-plane in type-IIA string theory. In
the limit the eleventh dimension opens up, the orientifold goes
over to the Kaluza-Klein monopole involving the transverse
Atiyah-Hitchin metric. In the same spirit, we expect that our
solutions describe  well-defined objects once embedded in
higher-dimensional gravity theories or string theories.

An issue is whether it is possible to create a pair of positive-
and negative-tension $q$-branes. The aforementioned vacuum
stability theorem is to ensure that such a process cannot possibly
take place. In the case of orientifolds, pair creation of
orientifold and anti-orientifold would be impossible simply
because the boundary condition at asymptotic infinity does not
match with that of the vacuum --- flat Minkowski space-time. In
the case of our solution, the asymptotic geometry at past or
future infinity reduces either to Milne Universe for $k=-1$ or to
degenerate space-time. Either way, they do not result in a flat
Minkowski space-time \emph{globally}. This implies that, even if we
may manufacture pair creation of positive and negative-tension
$q$-branes, the process does not lead to a vacuum instability
simply because the pair-creation geometry cannot be glued
smoothly to the flat Minkowski space-time. It is in this sense,
we believe, that the negative-tension objects inherent to our
solution leave the vacuum stability theorem unaffected even at
non-perturbative level.

\subsection{How singular is the time-like singularity?}
We now examine the behavior of waves near the time-like
singularity at $r=0$, and ask whether the singularity is
ameliorated when it is probed by waves rather than by
particles.\footnote{We limit our discussion to the massless case:
for the massive one, the singularity is already well behaved.}

This sort of the problem has been studied
previously~\cite{Marolf,Akihiro} in the context of static
space-times having time-like singularities. In some cases, it can
happen that space-times which appear singular when probed by
classical particles are not singular when these test particles
are treated quantum mechanically as waves. Qualitatively, this
occurs when an effective repulsive barrier is produced that does
not permit the particles to enter into the singularity, and
instead scatters them. More precisely, the singular region is not
singular to waves if these waves propagate through the
singularity in a definite and unique way. As explained
in~\cite{Akihiro}, mathematically, this condition is equivalent
to the condition that the time-translation operator for the waves
must be \emph{self-adjoint}. A sufficient condition to ensure this
property is if only one of the two linearly-independent solutions
to the equation
\begin{equation}\label{condition}
D^{\mu}D_\mu \psi \pm i \psi=0\,,
\end{equation}
is square-integrable.

In the present case, let us examine the solutions to the massless
Klein-Gordon equation  near the singularity  $r=0$, where the
equation becomes equivalent to eq.~(\ref{condition}). The condition
of non integrability of a solution translates into the following
condition on the wave function's radial part, $f(r)$:
\begin{equation}\label{connosi}
\parallel f \parallel^2 \propto \int_{0} dr r^{n} h_{+}h_{-}
\left( \frac{df}{dr} \right)^{2} \longrightarrow \infty \,,
\end{equation}
as $r \rightarrow 0$.

Since the Klein-Gordon equation reduces, for $r$ near 0, to:
\begin{eqnarray}
f''-\frac{(n-2)}{r}f'=0\,,
\end{eqnarray}
the two independent solutions to this equation behave as
\begin{eqnarray}
f(r) \sim c_0 + c_1 r^{n-1}\,,
\end{eqnarray}
for any dimension $n$, with arbitrary constants $c_0, c_1$. It is
clear that both of these solutions are normalizable, implying the
singularity is wave-singular.

\subsection{Temperature and entropy}
Given the explicit time dependence of the space-time in the
time-dependent regions I and III, one would expect particle
production takes place in these regions. This radiation would
indicate a quantum instability for the future region. A
calculation of this radiation is beyond the scope of the present
work, but we will make a preliminary analysis which shows that a
Hawking temperature can be associated with the static regions II
and IV of the space-time.

\paragraph{Hawking temperature.}

An indication that some observers may see excitations with a
thermal spectrum is offered by adopting the Hartle-Hawking
computation of the Hawking temperature for a
black-hole~\cite{NegMassPenrose}. These steps also lead to the
definition of a Hawking temperature for the space-time under
consideration, when applied to the static regions II and IV.

The estimate proceeds by performing a euclidean continuation of
the metric in this region by sending $t\rightarrow i\tau$, and
then demanding no conical singularity at the horizon in this
euclidean space-time. This condition requires the euclidean time
coordinate to be periodic $\tau \sim \tau + 2 \pi/\kappa$, and so
implicitly defines a temperature: $T = \kappa/(2\pi)$.

The $r$- and $\tau$-dependent parts of the euclidean metric in the
static region are:
\begin{eqnarray}
ds^2_E &=& |h_+|^{-1} h_-^{A+b-1} dr^2 + |h_+|
h_-^{A+1-(n-1)b} d\tau^2 \,, \nonumber \\
&\approx& h_-^{A+b-1} \left(\frac{r_+}{(n-1) \rho}\right) d\rho^2
+ h_-^{A+1 - (n-1)b} \left(\frac{(n-1) \rho }{r_+} \right)
d\tau^2\,, \nonumber\\
&\equiv& dR^2 + \kappa^2 \, R^2 d\tau^2 \,,
\label{Emet}
\end{eqnarray}
where $\rho = r_+ - r \ll r_+$ gives the coordinate distance from
the horizon. The last equality of eqs.~(\ref{Emet}) defines the
rescaled radial coordinate $R$ and the parameter
\begin{eqnarray}
\kappa = \frac{(n-1)}{2 r_+}  h_-^{1-nb/2} \,,
\end{eqnarray}
which determines the temperature. Here, $h_- = h_-(r_+) = |k -
(r_-/r_+)^{n-1}|$ denotes the value of this quantity at the
horizon.

We find in this way the temperature:
\begin{eqnarray}
T = \frac{\kappa}{2 \pi} = \frac{n-1}{4 \pi r_+}
\left|k -\left(\frac{r_-}{r_+} \right)^{n-1} \right|^{1 - nb/2}.
\end{eqnarray}
This reduces to previously obtained expressions for the special
cases where these metrics agree with those considered elsewhere.
In particular, it vanishes for extremal black-branes, for which $k
= 1$ and $r_- = r_+$. For the four-dimensional Schwarzschild-type
solution presented in section~\ref{Sec2}, we have $r_- = 0$ and so
$T = |k|(n-1)/(4 \pi r_+)$.

\paragraph{Entropy.}

The possibility of associating a temperature with a space-time
involving horizons immediately suggests that it may also be
possible to associate to it an entropy, using the thermodynamic
relation
\begin{eqnarray}
\frac{\partial S}{\partial (-M)} = \frac{1}{T} \,.
\end{eqnarray}
The unusual negative sign in this expression arises because of a
technical complication in defining an entropy in the present
instance. The complication arises because the entropy is
associated with degrees of freedom behind the horizon, where the
globally-defined time-like Killing vector changes direction. This
situation is very much like what happens for the de Sitter space,
for which the above expression is used to define the
entropy~\cite{GibbonsEntropy}.

For simplicity, we  restrict first  ourselves to the simplest
Schwarzschild solution with $r_- = 0$, $k=-1$ and $n=2$, for which
we have $T = 1/(4 \pi r_+)$ and $\mathcal{T} /V=M/V= -P/V_n = -
r_+/2GV_n$, with $V_n$ denoting the volume of the $n$-sphere. In
this case, the entropy becomes:
\begin{eqnarray}
\frac{\partial S}{\partial (-M)} = - 8 \pi GV_n \frac{M }{V}\,,
\end{eqnarray}
from which we integrate to find
\begin{eqnarray}
\frac{S}{V} = 4 \pi G V_n \left(\frac{M}{V} \right)^2,
\end{eqnarray}
where the integration constant is chosen to ensure $S(M=0) = 0$.

Note that, although both $S$ and $M$ both diverge due to  infinite
volume of the planar or the hyperbolic directions, the entropy and
tension per unit volume are finite, and are related in the same
way as are these quantities for a black-brane.  Notice also that
we retrieve the usual expression, $S = 4 \pi G M^2$ when we
specialize to the $k=1$ case of a black brane.

In the general case, the expression for the entropy will depend
on the electric charge as well. In order to extract the general
form of the entropy we follow the standard prescription in terms
of the euclidean action. Consider first the  definition of the
Gibbs free energy:
\begin{equation}\label{gibbs}
W=-T \log Z=T S+Q \Phi(r) - \mathcal{T}(r) \,,
\end{equation}
where $\mathcal{T} (r)$ is given in eq.~(\ref{tensione}), while $\Phi(r)$
is the potential associated with the $q+2$ form. Note that we
have here a sign change in the right-hand side of
eq.~(\ref{gibbs}) with  respect to the usual definition of the
free energy as explained above.

In the semiclassical approximation, we can identify the partition
function $Z$ with $e^{-I_{E}}$, where $I_{E}$ corresponds to the
euclidean action for the system  From this fact  we obtain
immediately
\begin{equation}\label{relationeucl1}
T S+Q \Phi(r) - \mathcal{T}(r)=T I_{E} \,.
\end{equation}
At this point, we need an expression for the euclidean action for
our system. This takes the form
\begin{equation}\label{euclidean}
I_{E}=-\int d^{d}x\sqrt{g}\left(\alpha
R-\lambda(\partial\phi)^{2}-\eta e^{-\sigma \phi} F^{2} \right)
-2\alpha \int d^{d-1}x \sqrt{h} K\,.
\end{equation}
where we have included the Gibbons-Hawking boundary action.

The contribution from a boundary at a surface of a fixed $r$ is
\begin{eqnarray}
\int d^{d-1}x \sqrt{h}K
&=&\frac{V(1-n)}{2T} \left[r_{-}^{n-1}-k\,r_{+}^{n-1}
+ (2+A-(n-1)b)r_-^{n-1}\left(\left(\frac{r_+}{r}\right)^{n-1}
- 1\right) \right], \nonumber \\
& = & \frac{\mathcal{T}}{4\alpha T}\,,\label{propT}
\end{eqnarray}
where  in the second equality we have used eq.~(\ref{tensione}).
Consider now the solutions that we have found for our system.
Following~\cite{GibbonsEntropy} we take the magnetic rather than
the electric solution, using the duality transformations given
in  section~\ref{sec3.1}. Substituting the solutions in
eq.~(\ref{euclidean}), it is straightforward to obtain a general
expression for the euclidean action in terms of the parameters of
the model. We then find the following expression that relates
eq.~(\ref{euclidean}) to other global quantities, and that allows
interesting manipulations of eq.~(\ref{relationeucl1}):
\begin{equation}\label{relationeucl2}
I_E=\frac{1}{2 T}\left(\mathcal{T}(r)- Q \Phi(r) \right).
\end{equation}
Substituting eq.~(\ref{relationeucl2}) into
eq.~(\ref{relationeucl1}), a simple calculation yields the general
relation
\begin{eqnarray}
S=-I_{E} \,.
\end{eqnarray}
At this point, we can write the general expression for the entropy
density $s$, using the known value of the temperature $T$, for
any curvature $k$. Indeed, for $k=1$, in which the entropy is
calculated outside the outer horizon, it is enough to change sign
on the last expression in eq.~(\ref{gibbs}). We obtain the
following compact form
\begin{equation}
s=\frac{S}{V} =\alpha 4\pi\,r_+^{n}\left| k-
\left(\frac{r_-}{r_+}\right)^{n-1}
\right|^{nb/2} = \frac{1}{4G} \sqrt{g_{nn}}\left|_{r_+} \right.,
\end{equation}
where in the last equation we have used  $\alpha= (16\pi G)^{-1}$.
It is remarkable that, for any $k$, the expression for the entropy
does not depend on the coordinate $r: g_{nn}$ corresponds to
the determinant of the induced metric on the $n$ spatial
dimensions, and it is calculated at the horizon $r_{+}$. In case
$k=1$, we obtain the well-known relation
\begin{equation}
S=\frac{\mathcal{A}}{4G}\,,
\end{equation}
where $\mathcal{A}$ is the area of the black-hole horizon. Again
for $k=-1$ and 0,  the area of the horizon is infinite but we can
still consider the entropy per unit volume. It is worth noting
that these quantities can be made finite by modding out the
planar or the hyperbolic subspace  by discrete subgroups of
$\ISO(n)$ and $\SO(n-1,1)$, respectively, as the operation would
leave the volume of the horizon finite.

\paragraph{Particle detectors.}

Where is the thermal distribution of particles which are described
by the temperature and entropy just defined? We propose an answer
for this which follows a similar discussion for de Sitter
space-time~\cite{BirrellDavies}. We follow here the argument as
described in ref.~\cite{stromDS}.

The key lies in the observation that the two-point propagator in
the static regions is periodic in imaginary time, as is seen by
the above derivation of the euclidean periodicity.  This implies
that the transition rate, $R(i \rightarrow j)$, for the
excitation of a simple particle detector from level $i$ to level
$j$ satisfies
\begin{equation}
R(i \rightarrow j) = R(j \rightarrow i)
e^{(E_j - E_i)/T}\,.
\end{equation}

This is precisely the relation which these rates must have if they
are to satisfy detailed balance in the presence of a thermal
distribution of particles. One thus infers that when it is in a
steady state, the detector responds as if it is in the presence of
such a thermal distribution.

\section{Discussion}\label{Sec6}

We now summarize our results, and indicate some directions for
future work.

\subsection{Summary}
In~\cite{gqtz}, a variety of solutions to the
$d\! =\! (n+q+2)$-dimensional dilaton, metric, antisymmetric-tensor
equations were found under the assumption that $q$-dimensional
subspace is flat and $n$-dimensional subspace is
maximally-symmetric with scalar curvature $kn(n-1)$, $k=0,\pm 1$.
Among the solutions, the $k=1$ class were identified to be the
black $q$-branes, first discussed in~\cite{gqtz}, but the $k=0,-1$
classes were new. One of the objectives in the present work was to
provide viable interpretations for the solutions belonging to the
new classes.

We interpreted that these solutions describe a non-trivial field
configuration produced by a pair of charged $q$-branes, whose
world surfaces are the time-like singularities of the metric. The
two $q$-branes carry equal but opposite-sign charges associated
with the antisymmetric tensor field. Tension of the two $q$-branes
are equal, but are generically negative-valued for both $k=-1$ and
$k=0$.

With such an interpretation, we have found that the metric near
the singularities is static (regions II and IV of
figure~\ref{fig1}), and describes the fields in the immediate
vicinity of the $q$-branes. By contrast, the metric at late times
becomes time-dependent, and we interpreted this as describing the
expansion of the metric as the two $q$-branes interact each other.
The metric must expand in this way because stability precludes
negative-tension branes from being free to move relative to the
other within a fixed background geometry. For $k=-1$, the metric
at asymptotically late times becomes flat. Region III of the
maximally-extended space-time describes the time-reversed process
of the situation in region I.

We also have shown that observers moving along time-like geodesics
can enter the static regions near the $q$-branes from the past
time-dependent region III, and can pass out from there into the
future time-dependent region I. These time-like geodesics are
repelled by the time-like singularities, and hence do not hit them
while evolving throughout the static regions.

The horizons of the space-time describe the set of events where
observers make the transition into or out of the static regions
near the $q$-branes. Before crossing the horizon out of the static
regions, there is only one $q$-brane inside an observer's past
light-cone. After crossing the horizon, however, both $q$-branes
are in the observer's past light-cone (but not in the future
light-cone).

The metric near the horizon  resembles that of the S-brane as
described in~\cite{stromin}. For the special case of 0-branes
$(q=0)$, our solutions agree precisely with the S-branes. However
the physical interpretation we are providing resembles more the
one of~\cite{cornalba} than that in~\cite{stromin}. The point is
that in the original S-brane solution the attention was
concentrated only on the time-dependent part of the geometry for
which an spacelike ``object'', the horizon, was assigned
physical properties such as charge. When considering the full
spacetime geometry it is then clear that the physical object
corresponds to the  singularity which being in the static region
allows for an unambiguous definition of charge and tension. The
singularities then are similar to orientifold
planes~\cite{cornalba} in the sense of having negative tension
and being constrained to a fixed point in the $r,t$ coordinates
$(r=0)$. In this sense they are free from potential instabilities
in the same way as the orientifold planes are. However orientifold
planes are well-defined objects in terms of a set of boundary
conditions, which our solutions are  not a priori constrained to
obey. Furthermore an orientifold plane usually preserves part of
supersymmetry. A preliminary study indicates that our solutions
are non supersymmetric. Therefore we believe that these
negative-tension branes represent a generalization of orientifold
planes.

Certain aspects of our solutions also share features with those
discussed recently in~\cite{cornalba, TDBackgrounds}, where
time-like orbifolds of flat space-time were considered and
potentially interesting cosmological evolution were obtained by
compactifying some extra dimensions. On the other hand, our
solutions do not have the problems recently raised for those
orbifold geometries (see for instance the last reference
in~\cite{TDBackgrounds}), since they are obtained by solving
Einstein's equations without the need of orbifold twists. In this
regard, our solutions can be considered as an explicit yet
relatively simple class of string theory backgrounds for
addressing various issues raised therein such as  stability,
causal structure and particle/string productions.

A feature of the solutions of~\cite{gqtz} is that they ought to be
ubiquitous to a wide class of supergravity, low-energy string
theories, and compactfications thereof. In case considered as
solutions of string theories,  they have well-defined domains of
validity which follow from the requirements of weak coupling
$(e^\phi \ll 1)$ and small curvature $(\alpha' \partial^2 \ll 1)$.
In that case, various stringy corrections are expected to alter
the regions close to the time-like singularities. On the other
hand, regions near the horizons would be free from such~corrections.

\subsection{Future directions} \label{future}
There are two important issues which we were able to resolve only
partially here. One is the question of the stability of these
space-times (for a recent discussion see~\cite{garysean}). We have
shown that the past horizons of these space-times are likely to be
unstable in precisely the same manner as the inner horizon of the
Reissner-Nordstr\"om black-hole. A more complete investigation of
stability is obviously of considerable interest. The second is the
question of quantum instability, and whether the time-dependent
fields in regions I and III give rise to particle production. We
have argued that there is a natural definition for the Hawking
temperature for the static space-times near the $q$-branes, and
this strongly suggests that this is associated with thermal
radiation as seen by the static observers. A more detailed
calculation of particle production is certainly desirable.

The time-dependent regions, I and III, of the space-times are also
of considerable interest, because they may open up a new avenue
for the cosmology of the early Universe. In the future
time-dependent region I, the space-time exhibits expansion of the
hyperbolic directions, and a past horizon without a past
space-like singularity. Since region III corresponds to the
time-reverse of region I, taken together the two
regions I and III offer an interesting realization of a
singularity-free cosmology, which bounces from a contracting to an
expanding Universe. One is eventually interested in a realistic
bouncing cosmology free from instabilities at the past horizon. A
possible prescription ensuring the stability would be a periodic
identification of the Killing coordinate, $r$, in the
time-dependent regions.

An obvious obstacle to constructing a realistic cosmology out of
the solutions eqs.~(\ref{tres}) and~(\ref{cuatro}) is that
co-moving observers do not see a homogeneous and isotropic space.
This objection needs not be fatal, as it may describe space-time
during the very early Universe --- perhaps during inflation --
before the Universe is really required to be isotropic and
homogeneous. Indeed, an attractive brane cosmology for these early
epoch has been proposed by utilizing brane-antibrane
interactions~\cite{BBInflation}. Alternatively, it may be the
higher-dimensional solutions which are of cosmological relevance.
After all, these space-times do have three-dimensional
hyper-surfaces which are homogeneous and isotropic. In the
specific metrics presented in~\cite{gqtz}, this usually requires
that the radial coordinate, $r$, describes a compact direction.
This would be problematic if the spacetime also includes the
static regions because, there, the $r$-coordinate would correspond
to a compact time direction. Having closed time-like curves, it
may also lead to orbifold instabilities~\cite{TDBackgrounds}.
Certainly, a direction for future work is to explore whether a
reasonable cosmology can be constructed free from these
pathologies.

We are currently investigating the above issues, and will report
results elsewhere.

\acknowledgments

We thank Stephon Alexander, Martin Bucher, Jim Cline, Miguel
Costa,  Damien Easson, Gary Gibbons, Christophe Grojean, Stefano
Kovacs, David Mateos, Rob Myers, Yunsoo Myung, Eric Poisson,
Ashoke Sen, Paul Townsend, Michele Trapletti, Neil Turok, Toby
Wiseman, Jung-Tay Yee, and Sunnyo Yi for helpful conversations.
F.Q, S.-J. R., G.T. and I.Z. thank ICTP, Trieste where part of
this project was developed. C.B. and S.-J.R. respectively thank
DAMTP and the Newton Institute for Mathematical Sciences at
Cambridge University for hospitality. C.B.'s research is
partially funded by grants from N.S.E.R.C. of Canada and F.C.A.R.
of Qu\'ebec. F.Q.'s research is partially funded by PPARC.
S.-J.R.'s research is partially funded by BK-21 Program through
SNU-Physics (Team 2), the KOSEF Interdisciplinary Research Grant,
the KOSEF Leading Scientist Program, and SNU-CNS Faculty Research
Award. I.Z.C. is supported by CONACyT (Mexico), Trinity College
(Cambridge) and Cambridge Overseas Trust. G.T. is partially
supported by the European TMR project ``Across the Energy
Frontier'' under contract HPRN-CT-2000-0148 and by the European
RTN project: Supersymmetry and the Early Universe" under contract
number HPRN-CT-2000-00152.

\end{document}